\documentclass{optica-article}

\usepackage{algorithm}
\usepackage{amsmath}
\usepackage{soul}
\usepackage{float}
\usepackage{upgreek}
\usepackage[noend]{algpseudocode}
\usepackage{bibunits}

\usepackage{hyperref}

\hypersetup{
    colorlinks=true,
    linkcolor=blue,
    filecolor=magenta,
    urlcolor=blue,
    pdftitle={Fast Nanophotonic Inverse Design using Precomputed Numerical Green Functions},
    pdfpagemode=FullScreen
}

\journal{opticajournal}
\articletype{Research Article}

\usepackage{lineno}

\begin{document}
\begin{bibunit}[opticajnl]

\title{Fast Nanophotonic Inverse Design using Precomputed Numerical Green Functions}

\author{Thi Phuong Thao Nguyen,\authormark{1,$\dag$} James Wang,\authormark{2,$\dag$} and Constantine Sideris\authormark{1,*}}

\address{\authormark{1}Department of Electrical Engineering, Stanford University, Stanford, CA 94305, USA\\
\authormark{2}Department of Electrical Engineering and Computer Sciences, University of California, Berkeley, CA 94720, USA\\
\authormark{$\dag$}The authors contributed equally to
this work.\\
\email{\authormark{*}sideris@stanford.edu}} 


\begin{abstract*} 
Inverse design has been transformative in nanophotonics, providing an automated means of realizing high-performance, non-intuitive devices. However, the primary bottleneck of existing inverse design approaches is their reliance on time-consuming and computationally expensive full-wave electromagnetic simulations. In this work, we extend the Precomputed Numerical Green Function (PNGF) method for the first time to nanophotonic inverse design, accelerating evaluation of the forward problem by multiple orders of magnitude with no loss in accuracy compared to the full-wave solution. After a single, fully parallelizable precomputation step, both the objective function and its gradient can be evaluated through simple calculations whose cost scales linearly with the size of the design region. A low-rank matrix update technique further reduces the cost of objective function evaluation, yielding sub-millisecond computation times per iteration. Design examples using both gradient-based level-set optimization and tile-flipping direct binary search often achieve more than three orders of magnitude speedup in design time, resulting in an ultrafast photonic inverse design platform that can design new devices from scratch in seconds.

\end{abstract*}

\begin{abstract*} 
Inverse design has been transformative in nanophotonics, providing an automated means of realizing high-performance, non-intuitive devices. However, the primary bottleneck of existing inverse design approaches is their reliance on time-consuming and computationally expensive full-wave electromagnetic simulations. In this work, we extend the Precomputed Numerical Green Function (PNGF) method for the first time to nanophotonic inverse design, accelerating evaluation of the forward problem by multiple orders of magnitude with no loss in accuracy compared to the full-wave solution. After a single, fully parallelizable precomputation step, both the objective function and its gradient can be evaluated through simple calculations whose cost scales linearly with the size of the design region. A low-rank matrix update technique further reduces the cost of objective function evaluation, yielding sub-millisecond computation times per iteration. Design examples using both gradient-based level-set optimization and tile-flipping direct binary search often achieve more than three orders of magnitude speedup in design time, resulting in an ultrafast photonic inverse design platform.

\end{abstract*}

\section{Introduction}
Integrated silicon photonics is a rapidly advancing technology platform with demonstrated potential to enable a wide range of applications, including telecommunications\cite{Tele}, quantum communication\cite{Quantum}, biochemical sensing\cite{Sensing}, large-scale LiDAR\cite{Lidar}, and AI hardware\cite{ONN}. These applications rely on photonic devices that guide and manipulate light as desired within a compact footprint. Early photonic devices were developed through intuition-driven, geometrically restricted design and empirical simulation studies, often resulting in bulky, suboptimal structures. Inverse design \cite{Molesky2018} has emerged as an effective alternative, enabling efficient exploration of highly non-intuitive designs within a parameterized design region and often yielding superior performance compared to conventional trial-and-error methods. In the inverse design paradigm, the user specifies the desired behavior, performance, and design constraints, and optimization algorithms coupled with electromagnetic solvers automate the discovery of structures that meet these specifications.

Gradient-free optimization methods, such as Direct Binary Search (DBS) \cite{Shen_Wang_Polson_Menon_2015,DBS_1,DBS_2}, Particle Swarm Optimization (PSO)\cite{BPS_1,BPS_2}, and Genetic Algorithms (GA)\cite{GA}, do not require derivative information with respect to the design variables. However, such algorithms typically require a large number of iterations to converge, particularly for phase-sensitive devices or large-scale, high-resolution structures, resulting in high overall computational cost. Several works have accelerated DBS by incorporating neural networks and statistical models \cite{DBS_NN,DBS_Bayesian} or by guiding the optimization with gradient information\cite{DBS_Gradient}. In contrast, gradient-based approaches leverage continuous relaxations and an additional adjoint simulation \cite{Adj_1,Adj_2,Adj_3,Adj_4,Adj_5,Adj_6} to compute the gradient efficiently, enabling rapid convergence and the exploration of complex, non-intuitive designs. However, continuous relaxation leads to ``grayscale'' designs containing intermediate material states that are not physically realizable, and additional binarization regularization or projection steps are required to recover binary, fabricable structures. The level-set method addresses this challenge by representing the structure as the zero contour of a continuous function, often constructed using radial basis functions \cite{Luo2009-sh,Wang2006}, whose boundary evolves incrementally according to the Hamilton--Jacobi equation \cite{Barth1998}. This yields smooth, well-defined boundaries that can be optimized with efficient gradient-based methods while maintaining a binary material distribution throughout the entire optimization.

Nevertheless, the key limitation of both gradient-based and heuristic optimization methods is the large number of full-wave simulations required to reach convergence, including gradient evaluations for gradient-based methods and repeated objective function evaluations for both classes of approaches. This issue is especially critical for 3D designs or large-scale structures with numerous design parameters, where each simulation is computationally expensive and many iterations may be required. An alternative to the numerical optimization methods discussed above is training machine learning (ML) and deep learning (DL) models \cite{NN_1,NN_2,NN_3,NN_4}, which replace full-wave simulations during the inverse design process with comparatively inexpensive model inference. However, preparing a high-quality training dataset typically requires a large number of full-wave simulations to sufficiently cover the design space, resulting in an often prohibitively high training overhead. Moreover, the trained models may exhibit limited accuracy and/or suffer from ``hallucinations,'' especially in regions of the design space that were underrepresented in the training data. Overall, a large number of full-wave simulations is still required to construct all of the aforementioned frameworks. Each of these simulations can take minutes, or even hours for 3D problems, even with commercially available software such as Lumerical or advanced solvers such as Meep \cite{Chen2025,MeepLicense,Hammond2022} and Augmented Partial Factorization (APF) \cite{Li2024,Li2024MultiChannel}. In addition, the simulation domain often comprises substantial static regions, such as the air box, input/output waveguides, and perfectly matched layers, which, despite never changing during the optimization, significantly increase the size of the simulation domain and the corresponding computational cost.

To take advantage of the fact that only the design region varies while large portions of the domain remain unchanged throughout the optimization, previous works have explored using the Schur complement~\cite{LR_2} and low-rank updates~\cite{LR_1}, exploiting the lower dimensionality of the output channels and of the design region compared to the full simulation domain. Most recently, the Precomputed Numerical Green Function (PNGF) method \cite{sun2026PNGF} was introduced for the inverse design of microwave electromagnetic devices, achieving near real-time optimization with DBS. In this paper, motivated by these developments, we extend the PNGF method for the first time to the inverse design of nanophotonic devices, demonstrating it with both the gradient-free DBS approach of \cite{sun2026PNGF} and a gradient-based level-set method with greedy updates. The method first performs a precomputation step to obtain the numerical Green functions of the design environment, which reduces the size of the formulation from the full simulation domain to the design region alone. Candidate designs within the optimization loop are then evaluated using low-rank matrix updates, further reducing the cost of each evaluation. Notably, both the figure of merit and the gradient are obtained through computations that scale linearly with the size of the design region, eliminating the need for full-wave or adjoint simulations without compromising accuracy. This formalism is implemented within a finite-difference frequency-domain (FDFD) framework and demonstrated using (i) a modified DBS algorithm that takes connectivity into account to improve fabricability, illustrating the strength of PNGF for small, incremental updates, and (ii) a gradient-based level-set method, demonstrating the compatibility of PNGF with gradient-based inverse design. As design examples, we first inverse design a 2D TM-mode power splitter with a footprint of $ 2.4 \times 2.4~\upmu\text{m}^2$. Next, to demonstrate the scalability of the approach to large, many-wavelength problems, we also design a broadband directional coupler operating from $1490~\text{nm}$ to $1610~\text{nm}$ with a coupling region of $4 \times 1.52~\upmu\text{m}^2$. The approach achieves three to four orders of magnitude faster design times than conventional FD methods for the examples considered, reducing hours to days of computational effort down to seconds. All comparisons were run on the same server (dual AMD Epyc 7763 CPUs) using 128 cores unless stated otherwise.

\section{Fast Structure Evaluation using Precomputed Numerical Green Functions}
\label{sec:PNGF}

\begin{figure}[ht!]
\centering\includegraphics[width=13cm]{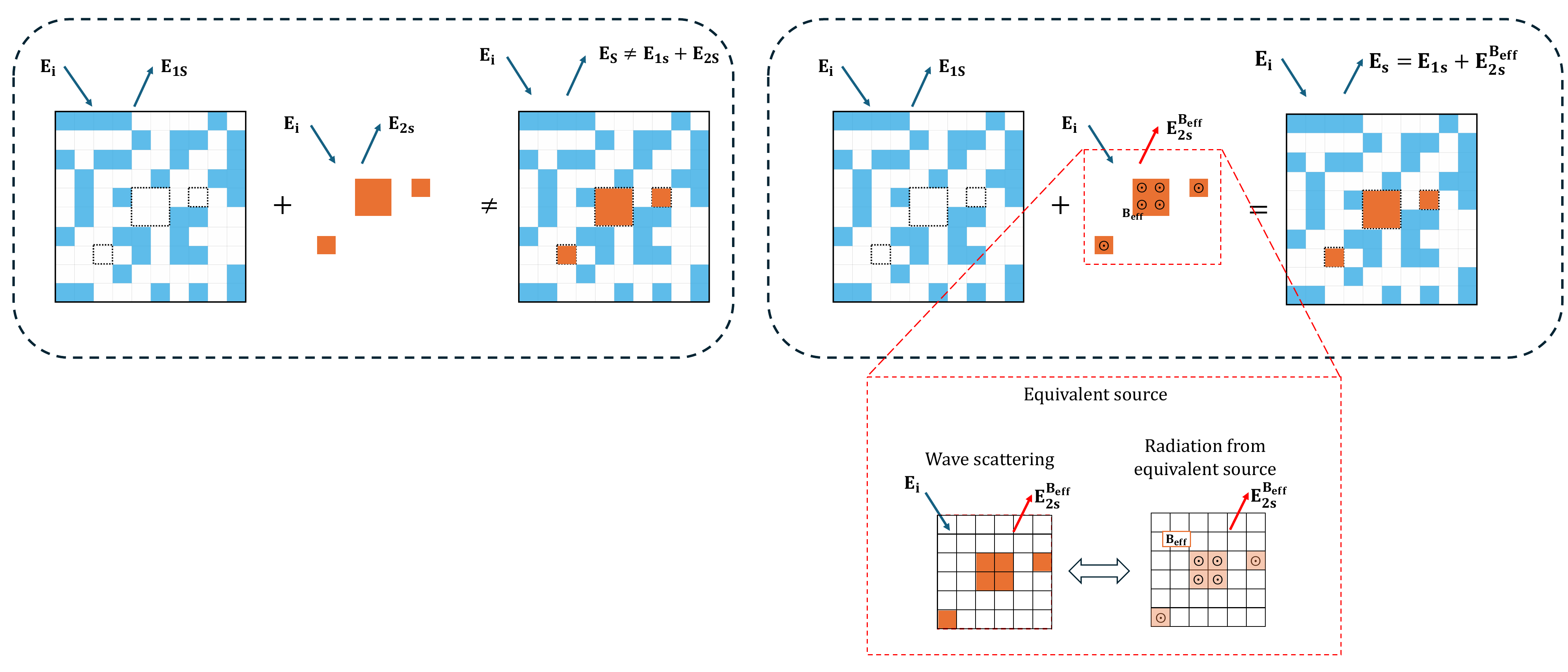}
\caption{\label{fig:eqv} Evaluation of a modified structure using source equivalence. Left: the field scattered by the modified structure cannot be obtained by superposing the fields scattered separately by the original structure and by the modification ($\mathbf{E_s} \neq \mathbf{E_{1s}} + \mathbf{E_{2s}}$), since this neglects the scattering interactions between the two. Right: the field of the modified structure \emph{is} obtained exactly by superposing the field of the original structure with the field $\mathbf{E^{B_{eff}}_{2s}}$ radiated by the equivalent sources $\mathbf{B_{eff}}$ that represent the modification. Inset: scattering of the incident wave by the modified pixels (left) is reproduced exactly by the radiation from the equivalent sources placed at those pixels (right).}
\end{figure}

In a typical inverse design workflow, the simulation domain can be separated into the design (dynamic) region, in which the design parameters ($\mathrm{\epsilon_r}$, $\mathrm{\mu_r}$) are allowed to change, and the non-design (static) region, which remains fixed throughout the optimization (e.g., input and output waveguides, insulator layers, surrounding free-space or dielectric regions, and absorbing boundary layers). The objective function may involve field quantities from both inside and outside the design region. In the traditional inverse design workflow, the entire domain is re-solved from scratch to recompute the fields and the corresponding objective function each time the design is modified, even when only small (i.e., sparse) modifications are made. This is computationally wasteful, particularly for large simulation domains in which only a small portion of the design region is modified at each iteration. The PNGF method instead treats each modification of the design region as a ``scatterer'' perturbation and computes the equivalent volumetric sources it induces, such that the new field is the superposition of the old field and the field radiated by these equivalent sources (Fig. \ref{fig:eqv}).

This section presents the method in three steps. First, we derive an exact update formula that evaluates the change in the objective function caused by an arbitrary modification of the design region using only a small set of precomputed matrices, whose dimensions are governed by the size of the design region rather than that of the full simulation domain. Second, we describe how these matrices are obtained in a single precomputation step. Finally, we show how the gradient of the objective function is obtained from the same precomputed quantities at negligible additional cost. We emphasize that PNGF makes no approximations and does not compromise accuracy: the effects of the static components are fully incorporated into the numerical Green functions during the precomputation phase, and every subsequent evaluation reproduces the full-wave solution exactly. As shown in Section \ref{sec:time}, as well as section \ref{SI:computation_cost} of the Supplementary Information, the computational cost of each evaluation scales linearly with the size of the design region, which is particularly impactful in practice, as the design region typically occupies only a small fraction of the overall simulation domain.

\subsection{Evaluation of Modified Structures via Low-Rank Matrix Updates}
\label{sec:lowrank}

In the finite-difference frequency-domain (FDFD) method, the electric field in a nonmagnetic medium is obtained by solving the linear system
\begin{equation}
\label{eq:FDFD}
    \mathbf{A}\mathbf{E} = \mathbf{B}, \qquad \mathbf{A} = \mathbf{C_h}\mathbf{C_e} - k_0^2\,\mathbf{diag}(\boldsymbol{\epsilon_r}),
\end{equation}
where $\mathbf{C_h}\mathbf{C_e}$ is the discretized curl-curl operator on the Yee grid \cite{Shin}, $\mathbf{diag}(\boldsymbol{\epsilon_r})$ is a diagonal matrix containing the relative permittivity at each of the $\mathrm{N}$ grid points, $k_0$ is the free-space wavenumber, and the $\mathrm{N_{inc}}$ columns of $\mathbf{B}=[\mathbf{b}_1,\mathbf{b}_2,\ldots, \mathbf{b}_{\mathrm{N_{inc}}}]$ are the source profiles of the excitations. The corresponding fields are the columns of $\mathbf{E} = \mathbf{A}^{-1}\mathbf{B}$. The complex quantities entering the objective function, such as mode-overlap amplitudes or field samples at the output ports, are linear functions of the fields and can be written as
\begin{equation}
\label{eq:objquantity}
    \mathbf{f} = \mathbf{S}^{\mathsf{T}}\mathbf{A}^{-1}\mathbf{B},
\end{equation}
where $\mathbf{S}$ is a tall $\mathrm{N} \times \mathrm{k}$ projection matrix whose $\mathrm{k}$ columns each encode one output quantity (e.g., a discretized mode-overlap integral; see Supplementary Information, section \ref{SI:mode_overlap}). A real scalar objective function, also called the figure of merit (FoM), $f_{\mathrm{FoM}}$, is then computed from $\mathbf{f}$, typically from the squared magnitudes of its entries.

\begin{figure}[ht!]
\hspace*{-1cm}
\centering\includegraphics[width=14cm]{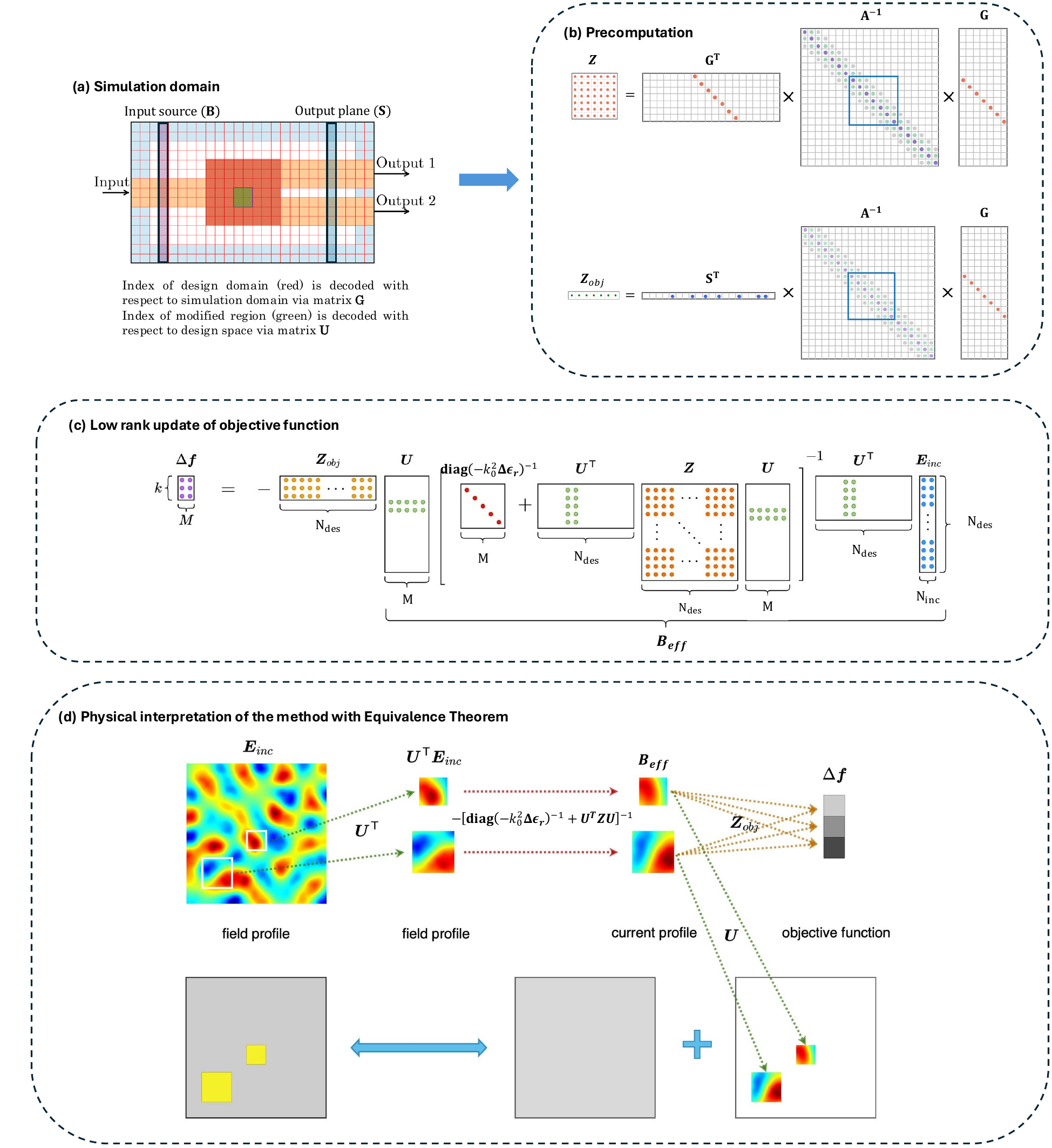}
\caption{\label{fig:PNGF_update} Overview of the PNGF method. (a) Simulation domain, comprising the static region with the input source ($\mathbf{B}$) and output plane ($\mathbf{S}$), the design region (red), whose grid indices are mapped to the simulation domain by the selection matrix $\mathbf{G}$, and the pixels modified at the current iteration (green), which are mapped to the design region by the selection matrix $\mathbf{U}$. (b) Precomputation of the numerical Green functions $\mathbf{Z} = \mathbf{G}^{\mathsf{T}}\mathbf{A}^{-1}\mathbf{G}$ and $\mathbf{Z_{obj}} = \mathbf{S}^{\mathsf{T}}\mathbf{A}^{-1}\mathbf{G}$. (c) Structure of the low-rank objective function update of Eq.~(\ref{eq:objupdate}), with the dimensions of each matrix indicated; the bracketed solve yields the equivalent sources $\mathbf{B_{eff}}$. (d) Physical interpretation via the equivalence theorem: the incident fields at the modified pixels, $\mathbf{U}^{\mathsf{T}}\mathbf{E_{inc}}$, determine the equivalent current sources $\mathbf{B_{eff}}$, which contribute to the objective function through $\mathbf{Z_{obj}}$.}
\end{figure}

The geometry of the problem is encoded in two selection matrices (Fig.~\ref{fig:PNGF_update}a): the tall $\mathrm{N} \times \mathrm{N}_{\mathrm{des}}$ matrix $\mathbf{G}$, which maps the $\mathrm{N}_{\mathrm{des}}$ grid points of the design region to their locations in the full simulation domain, and the tall $\mathrm{N}_{\mathrm{des}} \times \mathrm{M}$ matrix $\mathbf{U}$, which maps the $\mathrm{M} \le \mathrm{N}_{\mathrm{des}}$ design points modified at the current iteration of the optimization to their locations within the design region. Both are sparse 0--1 matrices satisfying $\mathbf{G}^{\mathsf{T}}\mathbf{G} = \mathbf{I}$ and $\mathbf{U}^{\mathsf{T}}\mathbf{U} = \mathbf{I}$, so that $\mathbf{G}^{\mathsf{T}}$ extracts the design region entries of a full-domain vector and $\mathbf{U}^{\mathsf{T}}$ extracts the modified entries of a design region vector.

Suppose the relative permittivity at the $\mathrm{M}$ selected points is changed by $\Delta\boldsymbol{\epsilon} = [\Delta\epsilon_1, \Delta\epsilon_2, \ldots, \Delta\epsilon_{\mathrm{M}}]$. Since the permittivity enters Eq.~(\ref{eq:FDFD}) only through the diagonal of $\mathbf{A}$, the modified system matrix is the rank-$\mathrm{M}$ perturbation
\begin{equation}
\label{eq:rankn}
    \mathbf{A}' = \mathbf{A} + \mathbf{G}\mathbf{U}\,\mathbf{D}\,\mathbf{U}^{\mathsf{T}}\mathbf{G}^{\mathsf{T}}, \qquad \mathbf{D} = \mathbf{diag}(-k_0^2\, \Delta\boldsymbol{\epsilon}).
\end{equation}
Applying the Woodbury matrix identity \cite{max1950inverting} to Eq.~(\ref{eq:rankn}) expresses the inverse of the modified wave operator in terms of the original inverse:
\begin{equation}
\label{eq:woodbury}
    \mathbf{A}'^{-1} = \mathbf{A}^{-1} -\mathbf{A}^{-1}\mathbf{G}\mathbf{U} \left[\mathbf{D}^{-1}+\mathbf{U}^\mathsf{T}\mathbf{G}^\mathsf{T}\mathbf{A}^{-1}\mathbf{G}\mathbf{U}\right]^{-1} \mathbf{U}^\mathsf{T}\mathbf{G}^\mathsf{T}\mathbf{A}^{-1}.
\end{equation}
Substituting Eq.~(\ref{eq:woodbury}) into Eq.~(\ref{eq:objquantity}) yields the exact change $\Delta\mathbf{f} = \mathbf{S}^{\mathsf{T}}(\mathbf{A}'^{-1} - \mathbf{A}^{-1})\mathbf{B}$ in the objective quantities caused by the modification:
\begin{equation}
\label{eq:objupdate}
    \Delta\mathbf{f} = -\mathbf{Z_{obj}}\mathbf{U} \left[\mathbf{D}^{-1}+\mathbf{U}^\mathsf{T}\mathbf{Z}\mathbf{U}\right]^{-1} \mathbf{U}^\mathsf{T}\mathbf{E_{inc}},
\end{equation}
where every occurrence of $\mathbf{A}^{-1}$ has been grouped into one of three matrices:
\begin{equation}
\label{eq:PNGFdef}
    \mathbf{Z} = \mathbf{G}^{\mathsf{T}}\mathbf{A}^{-1}\mathbf{G}, \qquad
    \mathbf{Z_{obj}} = \mathbf{S}^{\mathsf{T}}\mathbf{A}^{-1}\mathbf{G}, \qquad
    \mathbf{E_{inc}} = \mathbf{G}^{\mathsf{T}}\mathbf{A}^{-1}\mathbf{B}.
\end{equation}
These three matrices are the precomputed numerical Green functions of the method, and each has a clear physical interpretation (Fig.~\ref{fig:PNGF_update}b). The $\mathrm{N}_{\mathrm{des}} \times \mathrm{N}_{\mathrm{des}}$ matrix $\mathbf{Z}$ is the numerical Green functions of the design region: its $j$th column contains the field produced at every design point by a unit current source at the $j$th design point, with all interactions with the static region fully included. The $\mathrm{k} \times \mathrm{N}_{\mathrm{des}}$ matrix $\mathbf{Z_{obj}}$ maps unit currents in the design region directly to the objective quantities, and the $\mathrm{N}_{\mathrm{des}} \times \mathrm{N_{inc}}$ matrix $\mathbf{E_{inc}}$ contains the incident fields in the design region produced by the excitations $\mathbf{B}$. All three can be computed once for a given simulation environment, before the optimization begins, as described in Section~\ref{sec:precomp}. Thereafter, Eq.~(\ref{eq:objupdate}) evaluates any candidate modification exactly, without any additional full-wave simulation, at a cost that scales linearly with $\mathrm{N}_{\mathrm{des}}$ (Supplementary Information, section \ref{SI:computation_cost}).

Equation~(\ref{eq:objupdate}) is the algebraic form of the source equivalence picture of Fig.~\ref{fig:eqv}. By the volume equivalence and uniqueness theorems \cite{Volakis_Sertel_2012}, any dielectric structure can be replaced by a set of uniquely prescribed polarization currents that produce identical fields. The permittivity modification is therefore equivalent to the current sources $\mathbf{B_{eff}} = k_0^2\, \mathbf{diag}(\Delta\boldsymbol{\epsilon})\, \mathbf{U}^{\mathsf{T}}\mathbf{G}^{\mathsf{T}}\mathbf{E}'$ located at the $\mathrm{M}$ modified points, where $\mathbf{E}'$ is the total field of the modified structure. Since $\mathbf{E}'$ is not known a priori, $\mathbf{B_{eff}}$ is instead obtained directly by solving the $\mathrm{M} \times \mathrm{M}$ linear system appearing inside Eq.~(\ref{eq:objupdate}), which is a discretized form of the volume equivalence relation (see sections \ref{SI:eqv_theorem}-\ref{SI:compress_matrix}, of the Supplementary Information for the derivation):
\begin{equation}
\label{eq:eqvsource}
    \left[\mathbf{D}^{-1}+\mathbf{U}^\mathsf{T}\mathbf{Z}\mathbf{U}\right]\mathbf{B_{eff}} = -\mathbf{U}^\mathsf{T}\mathbf{E_{inc}}.
\end{equation}
The field scattered by the modification is then $\mathbf{E_s} = \mathbf{A}^{-1}\mathbf{G}\mathbf{U}\mathbf{B_{eff}}$, and its contribution to the objective quantities is $\Delta\mathbf{f} = \mathbf{S}^{\mathsf{T}}\mathbf{E_s} = \mathbf{Z_{obj}}\mathbf{U}\mathbf{B_{eff}}$, in agreement with Eq.~(\ref{eq:objupdate}) (Fig.~\ref{fig:PNGF_update}c,d).

When a candidate modification improves the FoM and is accepted, the three PNGF matrices must be updated so that they describe the new structure, which becomes the reference for subsequent iterations. Applying Eq.~(\ref{eq:woodbury}) to the definitions in Eq.~(\ref{eq:PNGFdef}) yields the update equations
\begin{align}
\label{eq:Eincupdate}
    \mathbf{E_{inc}} &\leftarrow \mathbf{E_{inc}} - \mathbf{Z}\mathbf{U}\left[\mathbf{D}^{-1}+\mathbf{U}^\mathsf{T}\mathbf{Z}\mathbf{U}\right]^{-1} \mathbf{U}^\mathsf{T}\mathbf{E_{inc}}, \\
\label{eq:Zobjupdate}
    \mathbf{Z_{obj}} &\leftarrow \mathbf{Z_{obj}} - \mathbf{Z_{obj}} \mathbf{U} \left[\mathbf{D}^{-1}+\mathbf{U}^\mathsf{T}\mathbf{Z}\mathbf{U}\right]^{-1} \mathbf{U}^\mathsf{T} \mathbf{Z}, \\
\label{eq:Zupdate}
    \mathbf{Z} &\leftarrow \mathbf{Z} - \mathbf{Z} \mathbf{U} \left[\mathbf{D}^{-1}+\mathbf{U}^\mathsf{T}\mathbf{Z}\mathbf{U}\right]^{-1} \mathbf{U}^\mathsf{T} \mathbf{Z},
\end{align}
where $\mathbf{Z}$ must be updated last, since the pre-update $\mathbf{Z}$ appears on the right-hand side of all three equations. Note that the update of $\mathbf{E_{inc}}$ is simply the design region restriction of the field superposition in Fig.~\ref{fig:eqv}, $\mathbf{E_{inc}} \leftarrow \mathbf{E_{inc}} + \mathbf{G}^{\mathsf{T}}\mathbf{E_s}$. The cost of an accepted update is dominated by Eq.~(\ref{eq:Zupdate}) and scales as $O(\mathrm{N}_{\mathrm{des}}^2)$ (Supplementary Information, section \ref{SI:computation_cost}). Since evaluations via Eq.~(\ref{eq:objupdate}) are far cheaper than updates, the method is especially efficient for optimizers that evaluate many candidate modifications but accept only a small fraction of them, such as DBS.

\subsection{Precomputation of the Numerical Green Functions}
\label{sec:precomp}

The most direct way to obtain $\mathbf{Z}$ and $\mathbf{Z_{obj}}$ is to excite a unit current source at each Yee grid point of the design region: projecting the resulting fields onto the design region yields the corresponding column of $\mathbf{Z}$, while projecting them onto the objective outputs via $\mathbf{S}^{\mathsf{T}}$ yields the corresponding column of $\mathbf{Z_{obj}}$. The required $\mathrm{N}_{\mathrm{des}}$ full-wave simulations are fully independent of one another and can therefore be run in parallel. For large numbers of design variables, however, a more efficient approach is to obtain both matrices in a single shot from the Schur complement produced by the augmented partial factorization (APF) technique \cite{APF}. To this end, an augmented matrix $\mathbf{P}$ is formed with $\mathbf{A}$ in its upper-left block and partially factorized as
\begin{equation}
\label{eq:APF}
\setlength{\arraycolsep}{2pt}
\renewcommand{\arraystretch}{1}
\mathbf{P} =
\begin{bmatrix}
\mathbf{A} & [\mathbf{G}\;\mathbf{0}] \\
\left[ \begin{matrix} \mathbf{G}^\mathsf{T} \\ \mathbf{S}^\mathsf{T} \end{matrix} \right] & \mathbf{0}
\end{bmatrix}
=
\begin{bmatrix}
\mathbf{P}_{L11} & \mathbf{0} \\
\mathbf{P}_{L21} & \mathbf{I}
\end{bmatrix}
\begin{bmatrix}
\mathbf{P}_{U11} & \mathbf{P}_{U12} \\
\mathbf{0} & -\mathbf{P}_{U22}
\end{bmatrix},
\end{equation}
where the zero-padding of the input block $[\mathbf{G}\;\mathbf{0}]$ keeps the augmented matrix square. The negative Schur complement $\mathbf{P_{U22}} =  \left[ \begin{smallmatrix} \mathbf{G}^\mathsf{T} \\ \mathbf{S}^\mathsf{T} \end{smallmatrix} \right] \mathbf{A}^{-1} [\mathbf{G}\;\mathbf{0}]$ is produced directly by the partial factorization, without ever forming $\mathbf{A}^{-1}$: $\mathbf{Z}$ is its top-left $\mathrm{N}_{\mathrm{des}}\times \mathrm{N}_{\mathrm{des}}$ block, and $\mathbf{Z_{obj}}$ is its bottom-left $\mathrm{k}\times \mathrm{N}_{\mathrm{des}}$ block. The incident-field matrix $\mathbf{E_{inc}}$ can be obtained from the same factorization by appending the source columns $\mathbf{B}$ to the input block. Notably, the computational cost of APF is nearly independent of the number of input and output channels \cite{APF}.

The precomputation needs to be performed only once for a given simulation environment and can be reused across optimization runs. For the examples presented in Section \ref{sec:DBS} and \ref{sec:LS}, the precomputation step took under 2 minutes for the power splitter and under 20 minutes for the broadband directional coupler.

\subsection{Gradient Calculation}
\label{sec:grad}
For gradient-based optimization, the gradient of the FoM is obtained from the same precomputed quantities at negligible additional cost. At any iteration, let $\boldsymbol{\epsilon_r}$ denote the current relative permittivity distribution inside the design region and define $\boldsymbol{\chi}=\boldsymbol{\epsilon_r}-\boldsymbol{\epsilon_r^0}$ as its deviation from the initial (precomputed) structure. The total derivative of the FoM with respect to the design variables follows from the chain rule:
\begin{equation}
\label{eq:gradtot}
    \frac{df_{\mathrm{FoM}}}{d\boldsymbol{\epsilon_r}} = \frac{\partial f_{\mathrm{FoM}}}{\partial \boldsymbol{\epsilon_r}} + 2\,\text{Re}\left[\frac{\partial f_{\mathrm{FoM}}}{\partial \mathbf{f}} \frac{\partial \mathbf{f}}{\partial \boldsymbol{\chi}}\right],
\end{equation}
where the derivatives with respect to the complex quantities $\mathbf{f}$ and $\boldsymbol{\chi}$ are Wirtinger derivatives, and $d\boldsymbol{\chi}/d\boldsymbol{\epsilon_r} = \mathbf{I}$ by definition. The first term of Eq.~(\ref{eq:gradtot}) is available analytically and typically arises from additional regularization terms, such as fabrication penalties or structural constraints. In the second term, $\partial f_{\mathrm{FoM}}/\partial \mathbf{f}$ is likewise known in closed form from the definition of the FoM, which is typically the squared norm of $\mathbf{f}$, as discussed in Section \ref{sec:lowrank}. The remaining factor $\partial \mathbf{f}/\partial \boldsymbol{\chi}$ follows from Eq.~(\ref{eq:objquantity}):
\begin{equation}
\label{eq:gradf}
    \frac{\partial \mathbf{f}}{\partial \boldsymbol \chi} = - \mathbf{S}^\mathsf{T} \mathbf{A}^{-1} \frac{\partial \mathbf{A}}{\partial \boldsymbol{\chi}} \mathbf{A}^{-1} \mathbf{B}.
\end{equation}
Since $\mathbf{A} = \mathbf{A_0} + \Delta\mathbf{A} = \mathbf{A_0} - k_0^2\, \mathbf{G}\,\mathbf{diag}(\boldsymbol \chi)\,\mathbf{G}^\mathsf{T}$, where $\mathbf{A_0}$ is the wave operator of the initial structure, we have $\partial\mathbf{A}/\partial\chi_i = -k_0^2\,\mathbf{G}\mathbf{X_i}\mathbf{G}^{\mathsf{T}}$, where $\mathbf{X_i}$ is a sparse matrix whose only nonzero entry is a ``1'' at position $(i,i)$. Evaluating Eq.~(\ref{eq:gradf}) element-wise then gives
\begin{equation}
\label{eq:gradelement}
    \frac{\partial \mathbf{f}}{\partial \chi_i } =  k_0^2\, \mathbf{S}^\mathsf{T} \mathbf{A}^{-1} \mathbf{G} \mathbf{X_i} \mathbf{G}^\mathsf{T} \mathbf{A}^{-1} \mathbf{B}
    =  k_0^2\, \mathbf{Z_{obj}} \mathbf{X_i} \mathbf{E_{inc}},
\end{equation}
which involves only the PNGF matrices of the current structure, maintained by the update equations~(\ref{eq:Eincupdate}) and (\ref{eq:Zobjupdate}). In other words, for the $i$th excitation, whose incident field is the column $\mathbf{e^i_{inc}}$ of $\mathbf{E_{inc}}$, and the $j$th objective output, observed on the row $\mathbf{z^j_{obj}}$ of $\mathbf{Z_{obj}}$, the gradient with respect to the permittivity of every design point is
\begin{equation}
\label{eq:gradsingle}
        \frac{\partial \mathbf{f}^{ij}}{\partial \boldsymbol{\epsilon_r}} =  k_0^2\, {\mathbf{z^j_{obj}}}^\mathsf{T} \odot  \mathbf{e^i_{inc}},
\end{equation}
where $\odot$ denotes the Hadamard (element-wise) product. The full gradient with respect to all $\mathrm{N}_{\mathrm{des}}$ design variables therefore costs only $O(\mathrm{N}_{\mathrm{des}})$ operations per excitation--objective pair, which is substantially cheaper than the additional adjoint simulation required in conventional gradient-based inverse design.

\section{Runtime Scaling of PNGF}
\label{sec:time}
\begin{figure}[ht!]
\centering\includegraphics[width=13.5cm]{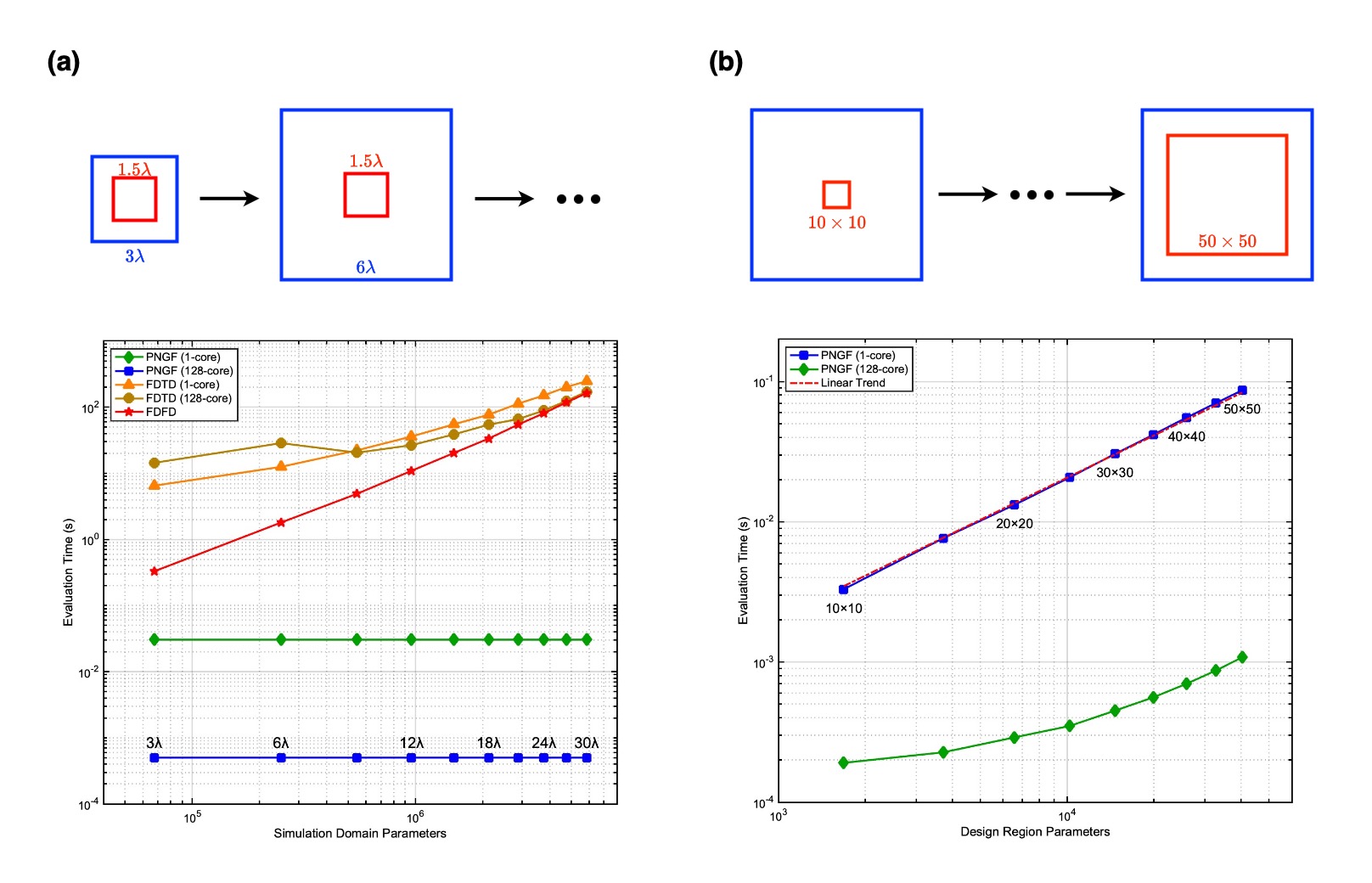}
\caption{\label{fig:speed} Scaling of the PNGF evaluation time with problem size. (a) Evaluation time versus the total number of simulation domain parameters for a fixed $1.5\lambda \times 1.5\lambda$ design region embedded in simulation domains ranging from $3\lambda \times 3\lambda$ to $30\lambda \times 30\lambda$. The PNGF evaluation time remains constant regardless of the size of the surrounding domain, whereas the runtimes of conventional FDFD and FDTD solvers grow with the domain size. (b) Evaluation time versus the number of design region parameters for a fixed simulation domain, with the design region growing from $10 \times 10$ to $50 \times 50$ tiles. The single-core PNGF evaluation time follows the expected linear trend (dashed line); the 128-core results appear sublinear for smaller problems, which do not fully utilize all available cores.}
\end{figure}

To analyze how the cost of the fast PNGF evaluation scales with problem size, we consider two scenarios (Fig.~\ref{fig:speed}). In the first, the size of the design region is fixed while the total simulation domain grows. Owing to the dimensionality reduction provided by PNGF, the evaluation time remains constant as the number of simulation parameters increases, whereas the runtime of conventional evaluation methods, such as FDTD or FDFD, scales with the size of the simulation domain. The speedup offered by PNGF is therefore even more significant when optimizing a structure embedded in a large simulation environment. In the second scenario, the simulation domain is fixed while the number of design parameters increases, and the PNGF evaluation time now scales with the number of design parameters. On a single core, the measured runtime is linear in the number of parameters, in agreement with the complexity analysis in Supplementary Information, section \ref{SI:computation_cost}. On multiple cores, the trend initially appears sublinear because smaller problems do not fully utilize the available parallelism. Together, these results show that PNGF is especially efficient for optimizing large structures and devices, with per-evaluation times in the millisecond range or below.

\section{Accelerated DBS using PNGF}
\label{sec:DBS}

DBS is a widely used optimization strategy for designing high-performance devices, owing to its simple implementation and the binary nature embedded in the method. The design region is divided into tiles, each of which can take one of a set of prescribed discrete levels (e.g., dielectric or air). During the optimization, one or several randomly chosen tiles are flipped at a time, and the objective function is re-evaluated. Based on the resulting change, the flip is either kept or reverted. The optimization has converged when no remaining flip improves the objective function. With the proposed method, the cost of evaluating a small per-iteration modification ($\mathrm{M} \ll \mathrm{N}_{\mathrm{des}}$) is essentially negligible, and the more expensive update of the PNGF matrices is performed only for accepted flips. Consequently, DBS emerges as a particularly well-suited optimization algorithm for PNGF. To show that the approach can be applied to a wide range of photonic devices, we first demonstrate the inverse design of a 50:50 power splitter operating at 1550 nm, followed by a compact broadband directional coupler that highlights the clear advantage of the method for multi-frequency, large-scale designs. Further examples of inverse design applied to other devices using the same strategy can be found in Supplementary Information, section \ref{SI:examples}. For the runtime comparison in this paper, the total optimization times of FDTD and FDFD are estimated by multiplying the runtime cost of a single simulation by the number of iterations required to reach convergence for each scenario.

\subsection{1550 nm 50:50 Power Splitter with Improved Connectivity}
\label{sec:PS_DBS}
\begin{figure}[ht!]
\centering\includegraphics[width=12cm]{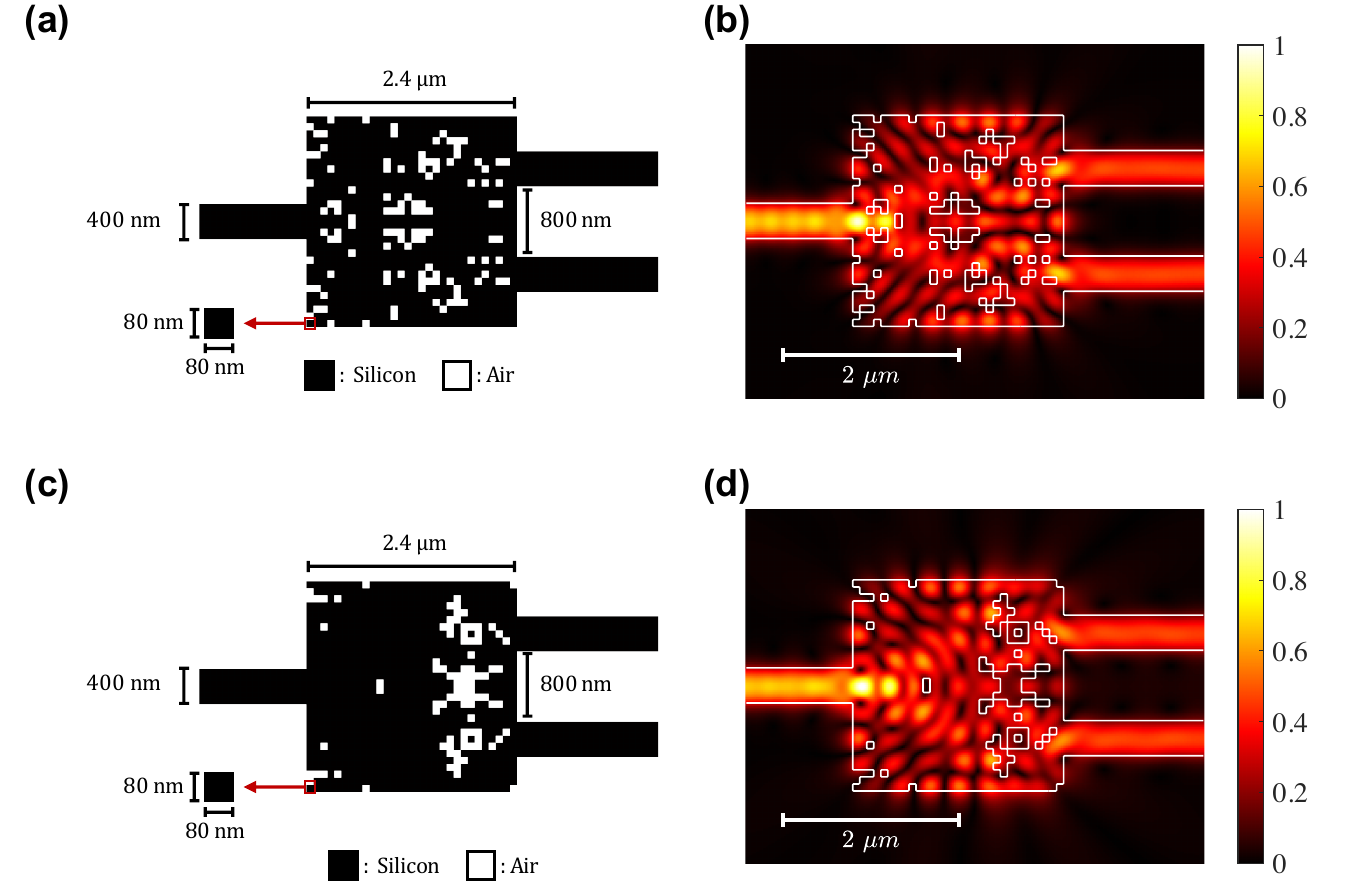}
\caption{Inverse-designed 50:50 power splitters at 1550 nm using standard and connectivity-enhanced DBS. (a) Optimized structure obtained with standard DBS: the $2.4 \times 2.4~\upmu\text{m}^2$ design region consists of $80 \times 80~\text{nm}^2$ silicon/air tiles and connects a 400-nm-wide input waveguide to two output waveguides separated by 800 nm. (b) Normalized $|E_z|$ distribution of the structure in (a) at 1550 nm, showing equal splitting between the two output ports; the white contour traces the silicon boundary. (c) Optimized structure obtained with the connectivity-enhanced (modified) DBS, exhibiting far fewer isolated single-tile features than (a). (d) Normalized $|E_z|$ distribution of the structure in (c).}
\label{fig:PS_DBS}
\end{figure}

A power splitter is a fundamental building block of photonic circuits, dividing the power carried by an incoming waveguide according to a predetermined ratio with low insertion loss. This enables controlled distribution of optical power for signal processing and interferometric functionalities. We begin with a 50:50 power splitter at 1550 nm, as it is one of the most widely used components. The parameterization for the DBS optimization is shown in Fig. \ref{fig:PS_DBS}a: a $2.4\, \upmu\mathrm{m} \times 2.4 \, \upmu\mathrm{m}$ region is allocated as the design region of the device and divided into $30 \times 30$ tiles of $80 \, \text{nm} \times 80\, \text{nm}$ each. Each tile takes one of two discrete states: 1, representing dielectric (silicon, $\epsilon_r = 11.7$), or 0, representing the background material (air, $\epsilon_r = 1$). Mirror symmetry is enforced about the $x$-axis. The precomputation step was completed in under two minutes for this power splitter and all devices in Supplementary Information.

We first attempt to maximize the transmission of the desired mode into the output channels. The optimization converges to a solution with total transmission $> 99.4\%$ ($> 49.7\%$ per channel) after $82$ accepted (``good'') flips and $1911$ rejected (``bad'') flips, taking a total of 12.1 seconds to run. The final structure and its electric field distribution are shown in Figs. \ref{fig:PS_DBS}a and \ref{fig:PS_DBS}b. However, the resulting structure exhibits poor connectivity between tiles, containing many isolated silicon ``islands'' and air ``holes'' that are undesirable for fabrication. To address this, we employ a modified DBS that prioritizes flipping tiles adjacent to existing features instead of selecting tiles fully at random; the details of the modified algorithm are provided in Supplementary Information, section \ref{SI:modified_DBS}. This approach promotes contiguous structures without significantly increasing the computational cost. The modified optimization converges after 87 good flips and 2906 bad flips, taking a total of 12.2 seconds, with a total transmission $> 99.1\%$. As shown in Fig. \ref{fig:PS_DBS}c, the final structure displays much better connectivity between regions compared to Fig. \ref{fig:PS_DBS}a. The same strategy extends beyond this device to arbitrary-ratio power splitters and other components, as further illustrated in Supplementary Information, section \ref{SI:examples}. The runtime performance of the method is summarized in Table \ref{table:PS_DBS}. 

\begin{table}[H]
\centering
\caption{\bf Convergence and Acceleration of Power Splitter using DBS}
\label{table:PS_DBS}
\resizebox{\textwidth}{!}{%
\begin{tabular}{cccccc}
\hline
 & FDTD (Lumerical) & FDFD & PNGF & Speed Up (FDTD) & Speed Up (FDFD) \\
\hline
DBS & 6.36 h & 0.8 h & 12.1 seconds & 1579$\times$ & 237$\times$\\
\hline
Modified DBS & 9.55 h & 1.2 h & 12.2 seconds & 2818$\times$ & 354$\times$\\
\hline
\end{tabular}
}
\end{table}

\subsection{Broadband Directional Coupler using PNGF-Accelerated DBS}
\label{sec:DC_DBS}

As shown in Sections \ref{sec:PNGF} and \ref{sec:time}, the complexity of the method is governed by the size of the design region and is independent of the size of the simulation domain. The advantage of PNGF over conventional full-wave-based optimization is therefore even more dramatic for large-scale, broadband problems. To illustrate this, we design a broadband directional coupler (DC) operating from $1490\ \text{nm}$ to $1610\ \text{nm}$. Beyond optical power distribution and signal routing, directional couplers are crucial components of various applications, such as interferometers, switches, and filters. However, a traditional directional coupler achieves its ideal splitting ratio at only a single frequency point and is considered narrowband, making its performance sensitive to signal distortion, temperature variations, and fabrication tolerances.

\begin{figure}[H]
\centering\includegraphics[width=12cm]{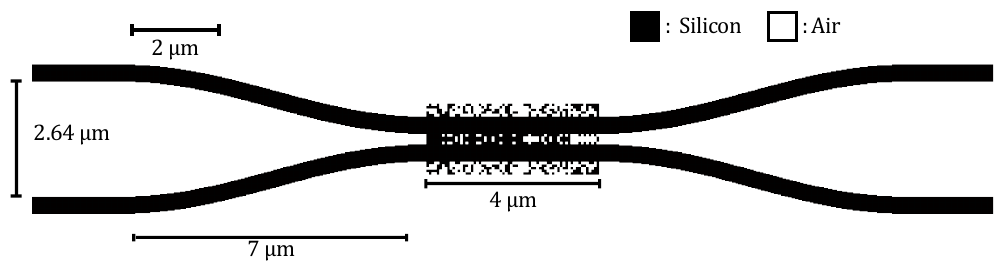}
\caption{Physical structure of the broadband directional coupler optimized using DBS. The pixelated $4 \ \upmu\mathrm{m}$ long coupling region (design region, $60 \times 60~\text{nm}^2$ tiles) connects two 400-nm-wide waveguides through $7 \ \upmu\mathrm{m}$ long $S$-bends; the input and output waveguides are initially separated by $2.64 \ \upmu\mathrm{m}$.}
\label{fig:DC_DBS_eps}
\end{figure}

The DC geometry is shown in Fig. \ref{fig:DC_DBS_eps}. Two $400 \ \text{nm}$ wide waveguides, initially separated by $2.64 \ \upmu\mathrm{m}$, are brought together by two $S$-bends of length $7 \ \upmu\mathrm{m}$ and height $1.2 \ \upmu\mathrm{m}$, chosen to minimize the loss introduced by the bends. These lead into a $4 \ \upmu\mathrm{m}$ long coupling region, in which the two waveguides are separated by $240 \ \text{nm}$ (edge to edge), followed by two identical $S$-bends to the output waveguides. The design problem consists of flipping $60 \ \text{nm} \times 60 \ \text{nm}$ tiles between silicon ($\epsilon_r = 11.7$) and background ($\epsilon_r = 1$). The design region spans the $3.96 \ \upmu\mathrm{m}$ length of the coupling region in the $x$-direction and extends in the $y$-direction from $300 \ \text{nm}$ below the lower waveguide to $300 \ \text{nm}$ above the upper waveguide, including the gap between the two waveguides. Mirror symmetry is again enforced about the $x$-axis. This design region admits $924$ design parameters and $2^{924}$ possible solutions. The objective function aggregates the power transmitted to the two output waveguides across the sampled wavelengths using harmonic means, with additional penalty terms enforcing even splitting and high total transmission:

\begin{equation}
\label{eq:DCobjective}
    \text{maximize:} \ f = \left[\sum_{i=1}^{n_{\lambda}} \left(\frac{1}{P_{i}^{up}}+\frac{1}{P_{i}^{down}}+\alpha\left(\frac{1}{P_{i}^{up}}-\frac{1}{P_{i}^{down}}\right)^2+\beta\,\frac{1}{P_{i}^{up}+P_{i}^{down}}\right)\right]^{-1}
\end{equation}
where $P_{i}^{up}$ and $P_{i}^{down}$ represent the normalized powers transmitted into the upper and lower output waveguides at the $i$th sampled wavelength. The wavelengths are sampled at Chebyshev nodes of the second kind with $n_{\lambda} = 7$, and we choose $\alpha = 8$ and $\beta = 5$ to balance the broadband and even-splitting objectives.The precomputation of the device at 7 frequencies took less than 11 minutes total.

\begin{figure}[H]
\centering\includegraphics[width=12cm]{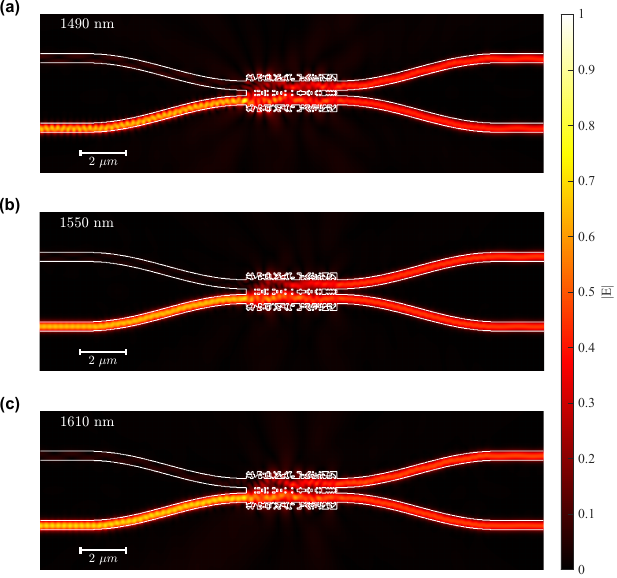}
\caption{Normalized $|E_z|$ distributions of the DBS-optimized directional coupler at (a) $1490\ \text{nm}$, (b) $1550\ \text{nm}$, and (c) $1610\ \text{nm}$, with the fundamental TM mode launched into the lower-left port. Nearly even splitting between the two output ports is maintained across the operating band. The white contours trace the silicon boundaries.}
\label{fig:DC_DBS_field}
\end{figure}

\begin{figure}[H]
\centering\includegraphics[width=12cm]{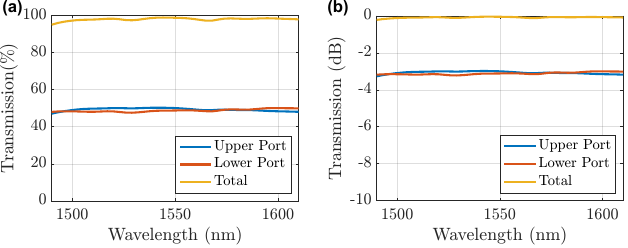}
\caption{Simulated broadband performance of the DBS-optimized directional coupler on (a) linear and (b) logarithmic (dB) scales, showing the power transmitted to the upper and lower output ports as well as the total transmission. Near-even splitting with $\approx 98\%$ total transmission is maintained across the entire $1490$--$1610 \ \text{nm}$ range.}
\label{fig:DC_DBS_performance}
\end{figure}

The design converges after $236$ good flips and $6423$ bad flips in $82.9$ seconds. The final physical structure, its electric field profiles, and its performance are shown in Figs. \ref{fig:DC_DBS_eps}, \ref{fig:DC_DBS_field}, and \ref{fig:DC_DBS_performance}, respectively. The average splitting ratio is $50.61 : 49.39$ with an average total transmission of $98.03\%$ over the $1490 \ \text{nm} - 1610 \ \text{nm}$ operating range. Running the same optimization with each iteration evaluated using the Lumerical FDTD solver on the same machine would take approximately 55 hours. The PNGF method therefore provides a speedup of more than $2400\times$ for this optimization (Table \ref{table:DC_DBS}).

\begin{table}[H]
\centering
\caption{\bf Convergence and Acceleration of Broadband Directional Coupler Optimization using DBS}
\label{table:DC_DBS}
\resizebox{\textwidth}{!}{%
\begin{tabular}{ccccc}
\hline
  FDTD (Lumerical) & FDFD & PNGF & Speed Up (FDTD) & Speed Up (FDFD) \\
\hline
 55.5 h & 51.8 h & 82.9 seconds &  2400$\times$ &  2236$\times$ \\
\hline
\end{tabular}
}
\end{table}

\section{Level-Set Method}
\label{sec:LS}
While DBS operates on a discrete tile grid, the PNGF framework is equally compatible with gradient-based optimization. To demonstrate this, we parameterize the structure of the devices using a continuous function $\phi(x,y)$, called the level-set function, whose zero contour $\phi(x,y)=0$ defines the boundary of the structure (illustrated in Fig. \ref{fig:PS_LS_rand}a):
\begin{equation}
\boldsymbol{\epsilon}(x,y) =
\begin{cases}
\epsilon_1 & \text{if } \phi(x,y) < 0 \\
\epsilon_2 & \text{if } \phi(x,y) > 0
\end{cases}
\end{equation}

The boundary then evolves incrementally along its normal direction following the Hamilton--Jacobi equation:

\begin{equation}
    \phi_n(x,y) \leftarrow \phi_{n-1}(x,y)+\alpha V_n |\nabla \phi(x,y)|,
\end{equation}
where $\alpha$ is the update step size and $V_n$ is the local boundary velocity.

The continuous level-set function is constructed from Radial Basis Functions (RBFs) centered at control knots placed on a coarse grid (Supplementary Information, section \ref{SI:level_set}). The permittivity on the level-set grid (Fig. \ref{fig:PS_LS_rand}b) is then obtained either by hard thresholding or by a smooth sigmoid projection, for example:
\begin{equation}
    \epsilon_{ls}(x,y)= \epsilon_1 + \frac{1}{2}(\epsilon_2-\epsilon_1) [ \tanh({\beta \phi(x,y)) }+1 ],
\end{equation}
and the permittivity profile on the simulation grid is obtained from the level-set grid by volumetric averaging (Fig. \ref{fig:PS_LS_rand}c). In addition, to avoid updating the entire design region at each iteration, a cutoff radius is introduced for each RBF, so that each knot update modifies only the pixels within the cutoff distance of that knot. This preserves the sparse-update structure that PNGF exploits.

Under typical conditions, the optimization can be started from a random initial guess and follow this workflow directly. However, the approach may suffer from near non-differentiability and can become trapped in local minima, particularly when initialized with an intuitive but suboptimal design, as discussed in sections \ref{SI:level_set} and \ref{SI:examples_levelset} of the Supplementary Information.

\subsection{50:50 Power Splitter}
We first apply this approach to the 50:50 power splitter, with mirror symmetry again enforced about the $x$-axis. Starting from a random initial guess, the RBF knots are placed uniformly on a grid coarser than the simulation grid, with spacing $dx_{\rho} = 0.04~\upmu\text{m}$ and cutoff radius $r_{cutoff} = 3~dx_{\rho}$. The structure is then optimized using greedy gradient descent with momentum, which improves convergence and reduces the tendency to become trapped in local minima.

\begin{figure}[h]
\centering
\includegraphics[width=\linewidth]{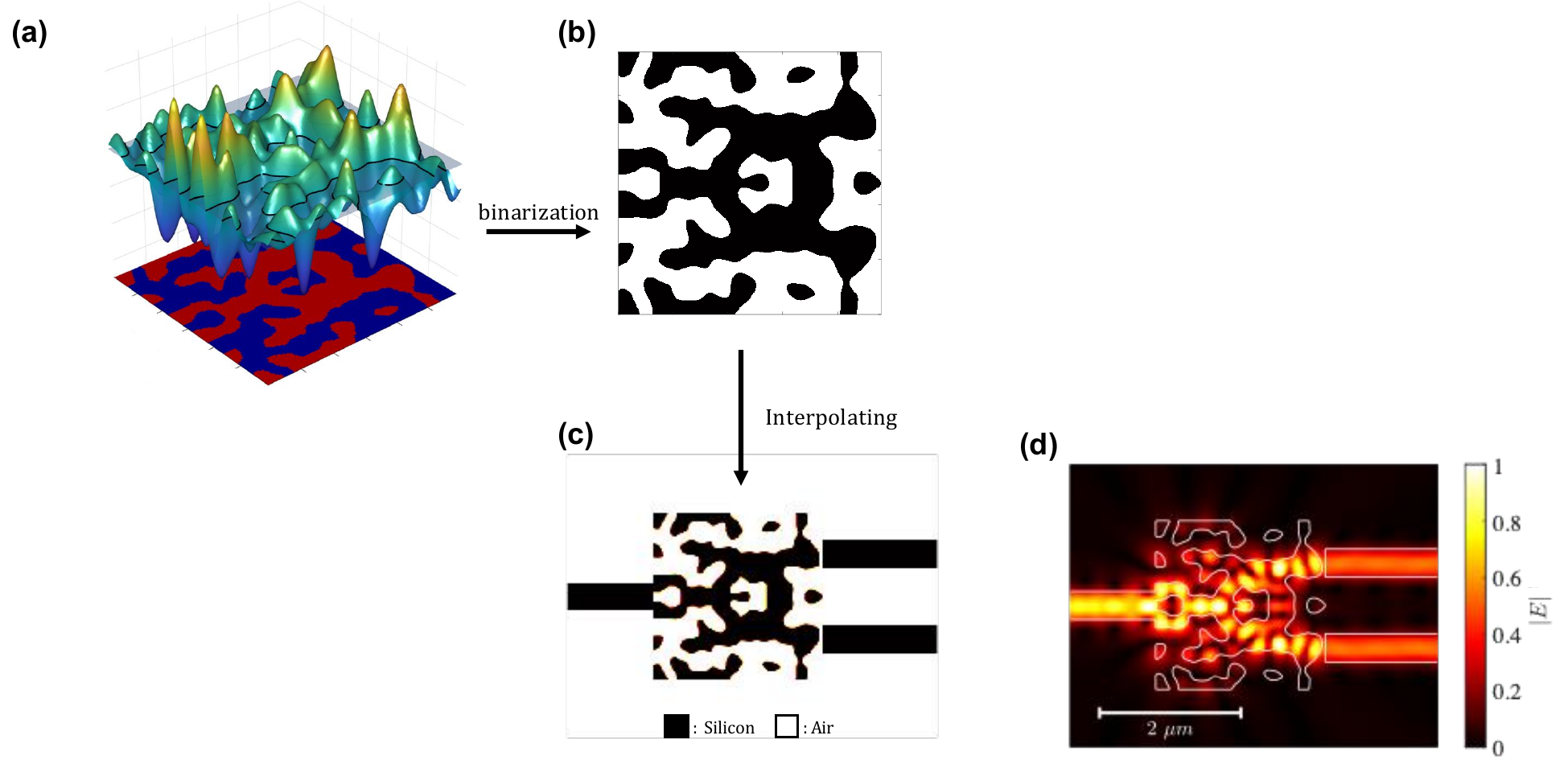}
\caption{\label{fig:PS_LS_rand} Level-set-based inverse design of the 50:50 power splitter. (a) Level-set function constructed from radial basis functions; its zero contour (gray plane) defines the boundary of the structure. (b) Binarized design obtained by projecting the level-set function onto the level-set grid. (c) Permittivity distribution on the simulation grid obtained by subpixel averaging; this is also the final optimized structure of the power splitter. (d) Normalized $|E_z|$ distribution of the final structure at 1550 nm; the white contour traces the silicon boundary.}
\end{figure}

The same objective function was used for evaluation as in Section \ref{sec:PS_DBS}. The optimization required 314 iterations and completed in 27.8 seconds, reaching a total transmission of $99.1 \%$ ($49.5 \%$ per port). The final design is shown in Fig. \ref{fig:PS_LS_rand}c, and the runtime benchmarks are summarized in Table \ref{table:PS_LS_rand}.

\begin{table}[H]
\centering
\caption{\bf Convergence and Acceleration of Power Splitter using level-set method}
\label{table:PS_LS_rand}
\resizebox{\textwidth}{!}{%
\begin{tabular}{ccccc}
\hline
 FDTD (Lumerical) & FDFD & PNGF & Speed Up (FDTD) & Speed Up (FDFD) \\
\hline
 2.0 h & 10.4 minutes & 27.8 seconds & 259$\times$ & 22$\times$\\
\hline
\end{tabular}
}
\end{table}

\subsection{Broadband Directional Coupler}
\begin{figure}[H]
\centering\includegraphics[width=12cm]{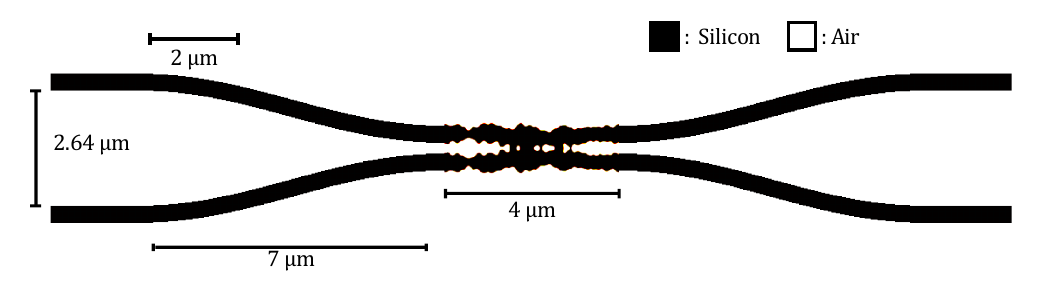}
\caption{Physical structure of the broadband directional coupler optimized using the level-set method. In contrast to the pixelated DBS design of Fig. \ref{fig:DC_DBS_eps}, the level-set parameterization yields a smooth, fabrication-friendly coupling region.}
\label{fig:DC_LS_eps}
\end{figure}

\begin{figure}[H]
\centering\includegraphics[width=12cm]{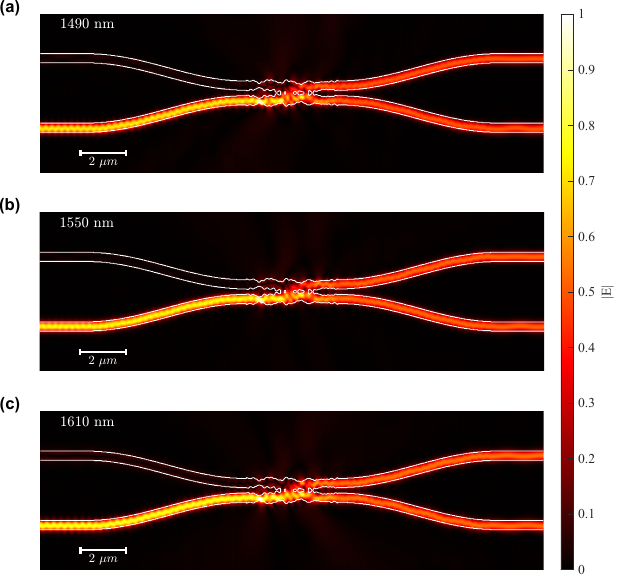}
\caption{Normalized $|E_z|$ distributions of the level-set-optimized directional coupler at (a) $1490\ \text{nm}$, (b) $1550\ \text{nm}$, and (c) $1610\ \text{nm}$, with the fundamental TM mode launched into the lower-left port; the white contours trace the silicon boundaries.}
\label{fig:DC_LS_field}
\end{figure}

\begin{figure}[H]
\centering\includegraphics[width=13cm]{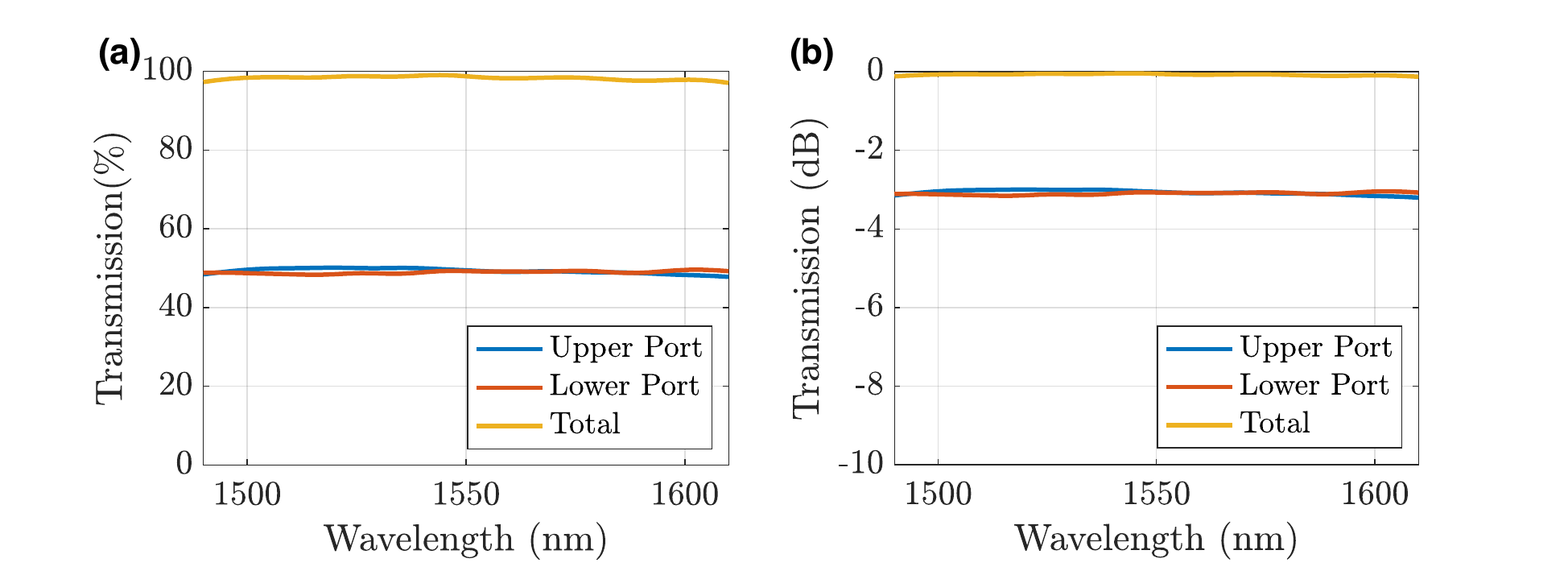}
\caption{Simulated broadband performance of the level-set-optimized directional coupler on (a) linear and (b) logarithmic (dB) scales, showing the power transmitted to the upper and lower output ports as well as the total transmission across the $1490$--$1610 \ \text{nm}$ operating range.}
\label{fig:DC_LS_performance}
\end{figure}

We now apply the PNGF-accelerated level-set framework to the broadband directional coupler. In this case, we inverse design the coupling region of the device, which has a footprint of $4 \times 1.52~\upmu\text{m}^2$. The RBF knots are placed uniformly with spacing $dx_{\rho} = 0.02~\upmu\text{m}$, spread $\sigma = 2dx_{\rho}$, and cutoff radius $r_{cutoff} = 3dx_{\rho}$. The sampled wavelengths and the objective function are the same as in Section \ref{sec:DC_DBS}. The optimization is initialized from a conventional directional-coupler-like junction comprising two closely spaced parallel waveguides separated by $0.24~\upmu\text{m}$. From this initial guess, the design converges within 367 iterations in under 5 minutes to the structure shown in Fig. \ref{fig:DC_LS_eps}, whose field distributions across the operating band are shown in Fig. \ref{fig:DC_LS_field}. The average total transmission across the operating band is 98.25$\%$ with an average splitting ratio of 50.4 : 49.6 (Fig. \ref{fig:DC_LS_performance}). Although this optimization requires more time than the DBS design of Section \ref{sec:DC_DBS}, it converges in significantly fewer iterations, produces a smooth, fabrication-friendly design, and still achieves speedups of at least $35\times$ compared to using conventional full-wave solvers (Table \ref{table:DC_LS}).

\begin{table}[H]
\centering
\caption{\bf Convergence and Acceleration of Broadband Directional Coupler Optimization using the Level-Set Method}
\label{table:DC_LS}
\resizebox{\textwidth}{!}{%
\begin{tabular}{ccccc}
\hline
 FDTD (Lumerical) & FDFD & PNGF & Speed Up (FDTD) & Speed Up (FDFD) \\
\hline
 3.1 h & 2.9 h & 4.9 minutes & 37$\times$ & 35$\times$\\
\hline
\end{tabular}
}
\end{table}

\section{Outlook}

This paper introduced the Precomputed Numerical Green Functions method for ultrafast simulation and optimization of nanophotonic structures. By reducing the computational dimensionality from the full simulation domain to the design region, the method achieves speedups of up to several orders of magnitude compared to conventional full-wave simulation without any compromise in accuracy. Since the method is derived directly from Maxwell's equations with no approximations, it can be applied to a wide range of photonic devices, as well as other electromagnetic devices in general.

From the perspective of optimization algorithms, any method that makes small incremental updates per iteration is well suited to PNGF due to the nature of the low-rank update algorithm. In the main text, we demonstrated PNGF with DBS and with a greedily updated level-set parameterization. However, we have included additional examples in the Supplementary Information, including a meta-atom (rotatable-cell) parameterization example in Supplementary Information, section \ref{supSI:RDBS} demonstrating that the method is also compatible with other parameterization strategies, such as grating-coupler structures and spline-based boundary parameterizations.

Future work will focus on extending the method to fully-vectorial three-dimensional problems. Since the computational cost depends solely on the size of the design region and the number of modified variables, rather than on the full simulation domain, the method is expected to provide even greater computational advantages in 3D, considering the correlation of the fields in all three directions. For large 3D problems, APF may become impractical for the precomputation step due to the significant memory required for the matrix factorization; however, this overhead can be mitigated using iterative methods such as the generalized minimal residual method (GMRES) \cite{Saad1986GMRES}, domain decomposition \cite{Skarda2022Metasurface,Lin2019OverlappingDomains}, or advanced preconditioning techniques \cite{Groth2019Circulant,Zhang2021FastPreconditioned}. In addition, principal component analysis \cite{10686031} can be employed to further enhance convergence efficiency and simulation accuracy and reduce the number of unknowns.

Finally, although demonstrated here using finite-difference schemes, the formalism is also compatible with other numerical methods, such as finite-element and integral-equation approaches: the same low-rank updates and parameterizations can be applied to any discretized form of the wave equation, and the corresponding precomputed quantities can be identified from the derivation in Section \ref{sec:PNGF}.

\begin{backmatter}
\bmsection{Acknowledgment}
The authors gratefully acknowledge support by the Air Force Office of Scientific Research (FA9550-26-1-B177) and the National Science Foundation (CCF-2047433).
\end{backmatter}

\putbib[sample]

\end{bibunit}


\begin{bibunit}[opticajnl]
\clearpage

\begingroup
\setlength{\parindent}{0pt}
\raggedright

{\sffamily\bfseries\fontsize{25}{28}\selectfont
Fast Nanophotonic Inverse Design
using Precomputed Numerical Green
Functions: supplemental document\par}

\vspace{0.8 cm}

{\normalsize
This document provides supporting information for
``Fast Nanophotonic Inverse Design using
Precomputed Numerical Green Functions.''\par}

\endgroup

\setcounter{section}{0}
\setcounter{subsection}{0}
\setcounter{figure}{0}
\setcounter{table}{0}
\setcounter{equation}{0}
\setcounter{algorithm}{0}

\renewcommand{\thesection}{S\arabic{section}}
\renewcommand{\thesubsection}{S\arabic{section}.\arabic{subsection}}
\renewcommand{\thefigure}{S\arabic{figure}}
\renewcommand{\thetable}{S\arabic{table}}
\renewcommand{\theequation}{S\arabic{equation}}
\renewcommand{\thealgorithm}{S\arabic{algorithm}}


\section{Equivalence Theorem}
\label{SI:eqv_theorem}
Considering the nonmagnetic case of Maxwell's equations, with no variation in the permeability profile ($\boldsymbol{\mu} = \mu_0$) and no magnetic current excitations ($\boldsymbol{M} = 0$), the electric field satisfies

\begin{equation}
    \nabla  \times \nabla  \times \boldsymbol{E} \ -  \ \omega^2\mu_0\epsilon_0\boldsymbol{\epsilon}\boldsymbol{E} \ = -i\omega\mu_0\boldsymbol{J}
\end{equation}

Any scatterer ($\boldsymbol{\epsilon} \neq \boldsymbol{I}$) can be replaced with an equivalent polarization current $\boldsymbol{J_P} = i\omega\epsilon_0(\boldsymbol{\epsilon} - \boldsymbol{I})\boldsymbol{E}$, where $\boldsymbol{\epsilon}$ is the relative permittivity distribution of the scatterers and $\boldsymbol{I}$ is equal to one everywhere. After determining the polarization current, the equation above can be transformed into the free-space Maxwell's equations:

\begin{equation}
\label{eq:tot}
    \nabla  \times \nabla  \times \boldsymbol{E} \ -  \ \omega^2\mu_0\epsilon_0\boldsymbol{E} \ = -i\omega\mu_0(\boldsymbol{J+J_P})
\end{equation}

With this expression, the total field $\boldsymbol{E}$ separates into the incident field $\boldsymbol{E_i}$, generated by the excitation $\boldsymbol{J}$, and the scattered field $\boldsymbol{E_s}$, generated by the polarization current $\boldsymbol{J_P}$. To express $\boldsymbol{E_s}$ in terms of $\boldsymbol{J_P}$, we consider the electric field produced by an arbitrary volumetric electric current density distribution in free space:

\begin{equation}
    \boldsymbol{E} = \boldsymbol{E_i} - \frac{1}{i\omega \epsilon_0}\int_V{\boldsymbol{G_0}(\boldsymbol{r,\boldsymbol{r'}})\ \boldsymbol{J_P}(\boldsymbol{r'})\ d\boldsymbol{V}}
\end{equation}
where $\boldsymbol{G_0}$ is dyadic free-space Green functions.

Substituting the expression $\boldsymbol{E} = \frac{1}{i\omega\epsilon_0}(\boldsymbol{\epsilon}-\boldsymbol{I})^{-1}\boldsymbol{J_P}$, the polarization current density distribution can be found without prior knowledge of the total field by solving the following integral equation:

\begin{equation}
\label{eq:eqvint}
    \boldsymbol{J_P} + (\boldsymbol{\epsilon}-\boldsymbol{I})\int_V{\boldsymbol{G_0}(\boldsymbol{r,\boldsymbol{r'}})\ \boldsymbol{J_P}\ d\boldsymbol{V}} = i\omega\epsilon_0(\boldsymbol{\epsilon}-\boldsymbol{I})\boldsymbol{E_i}
\end{equation}

Notice that this integral equation solves for the polarization current density collocated with the scatterers and yields zero in the free-space region, as expected. The equation holds in general for a continuous distribution of $\boldsymbol{\epsilon}$.

\section{Discretized Equivalence Theorem}
\label{SI:discretized_eqv_theorem}
In a discretized scheme with a nonmagnetic formulation ($\mu = \mu_0$), the total electric field is obtained by solving the linear system $\mathbf{AE}=\mathbf{B}$, where $\mathbf{A}$ is the discretized wave operator and $\mathbf{B}=-j\omega \mu_0 \mathbf{J}$ is the source profile. In FDFD, $\mathbf{A} = \nabla\times\nabla\times -k_0^2 \boldsymbol{\epsilon}$, where $\boldsymbol{\epsilon}$ is a diagonal matrix storing the relative permittivity profile of the simulation domain. Assume that at some point during the optimization the relative permittivity profile differs from its initial state $\boldsymbol{\epsilon_{bg}}$ by $\boldsymbol{\Delta \epsilon_r}=\boldsymbol{\epsilon_r}-\boldsymbol{\epsilon_{bg}}$. Throughout the following derivation, $\mathbf{A}$ denotes the initial FDFD operator constructed from $\boldsymbol{\epsilon_{bg}}$ and is not updated to include this permittivity change. The discrete counterpart of the volume equivalence theorem is $\mathbf{B_p}=k_0^2\Delta\boldsymbol{\epsilon_r}\mathbf{E}$, and hence the discrete counterpart of Eq. (\ref{eq:tot}) is:

\begin{equation}
    \mathbf{A}\mathbf{E} = (\mathbf{B+B_p}) \\
    \rightarrow \mathbf{E} = \mathbf{E_{i}} +\mathbf{A}^{-1} \mathbf{B_{p}} \rightarrow \mathbf{B_p} - k_0^2\Delta\boldsymbol{\epsilon_r} \mathbf{A}^{-1} \mathbf{B_{p}}= k_0^2\Delta\boldsymbol{\epsilon_r} \mathbf{E_i} 
\end{equation}

This is the discrete counterpart of Eq. (\ref{eq:eqvint}):
\begin{equation}
    \left[ \mathbf{I} - k_0^2 \boldsymbol{\Delta} {\epsilon_r} \mathbf{A}^{-1} \right] \mathbf{B_{p}} = k_0^2\Delta\boldsymbol{\epsilon_r} \mathbf{E_i} 
\end{equation}

To simplify the calculation and reduce its cost, we multiply both sides of the equation by $-[k_0^2\Delta\boldsymbol{\epsilon_r}]^{-1} = -\mathbf{diag}(k_0^{-2}\Delta{\epsilon_i}^{-1})$, which gives:

\begin{equation}
\label{eq:diseqv}
    \left[ -[k_0^2\Delta\boldsymbol{\epsilon_r}]^{-1} +  \mathbf{A}^{-1} \right] \mathbf{B_{p}} =  -\mathbf{E_i} 
\end{equation}

\section{Selection Matrices for Encoding Modifications}
\label{SI:compress_matrix}
Let $\mathbf{G}$ and $\mathbf{U}$ denote semi-unitary binary matrices of size $N \times N_{des}$ and $N_{des}\times M$, respectively. These are the masking (selection) matrices that select the design region within the simulation domain and the modified pixels within the design region. They satisfy $\mathbf{G}^\mathsf{T}\mathbf{G} = \mathbf{I_{N_{des}}}$ and $\mathbf{U}^\mathsf{T}\mathbf{U} = \mathbf{I_{M}}$. Considering a modification encoded by $\mathbf{U}$ on a set of $M$ pixels, $\Delta\boldsymbol{\epsilon} = [\Delta\boldsymbol{\epsilon}_1, \Delta\boldsymbol{\epsilon_2}, ...\Delta\boldsymbol{\epsilon_M}]$, the source profile $\mathbf{B_{eff}}$ generated by these $M$ pixels satisfies:

\begin{equation}
\begin{gathered}
    \label{eq:diseqvGU}
     -\left[ -[k_0^2\Delta\boldsymbol{\epsilon_r}]^{-1} +  \mathbf{A}^{-1} \right] \mathbf{(GU)}\mathbf{B_{eff}} = \mathbf{E_i}
\end{gathered}
\end{equation}

Since these equivalent sources appear only at the modified pixels, Eq. (\ref{eq:diseqvGU}) can be projected onto the modified pixels and written as:

\begin{equation}
\begin{gathered}
    \label{eq:diseqvU}
     - \left[ \mathbf{diag}(-k_0^2 \mathbf{\Delta}\boldsymbol{\epsilon})^{-1}+  \mathbf{(GU)}^\mathsf{T}\mathbf{A}^{-1} \mathbf{(GU)} \right] \mathbf{B_{eff}} = \mathbf{(GU)}^\mathsf{T}  \mathbf{E_i}
\end{gathered}
\end{equation}

Denoting $\mathbf{E_{inc}}=\mathbf{G}^\mathsf{T} \mathbf{E_i}$ and $\mathbf{Z} = \mathbf{G^\mathsf{T}A^{-1}G}$, this becomes:

\begin{equation}
\label{SIeq:eqvsource}
    \left[\mathbf{diag}(-k_0^2 \mathbf{\Delta}\boldsymbol{\epsilon})^{-1}+\mathbf{U}^\mathsf{T}\mathbf{Z}\mathbf{U}\right]\mathbf{B_{eff}} = -\mathbf{U}^\mathsf{T}\mathbf{E_{inc}}
\end{equation}

which is the equivalent-source system used in the main text.

\section{Woodbury Matrix Identity}
\label{SI:Woodbury}
In linear algebra, the Woodbury matrix identity states that the inverse of a rank-$\mathrm{k}$ update to a matrix (i.e., the addition of a rank-$\mathrm{k}$ matrix to the original matrix) can be computed by applying a corresponding rank-$\mathrm{k}$ update to the inverse of the original matrix. The identity is expressed mathematically as:

\begin{equation}
    (\mathbf{A} + \mathbf{B}\mathbf{C}\mathbf{D})^{-1} = \mathbf{A}^{-1} - \mathbf{A}^{-1}\mathbf{B}(\mathbf{C}^{-1}+\mathbf{D}\mathbf{A}^{-1}\mathbf{B})^{-1}\mathbf{D}\mathbf{A}^{-1}
\end{equation}

where $\mathbf{A}$ is an $\mathrm{n} \times \mathrm{n}$ invertible matrix, $\mathbf{B}$ is an $\mathrm{n} \times \mathrm{k}$ matrix, $\mathbf{C}$ is an $\mathrm{k} \times \mathrm{k}$ invertible matrix, and $\mathbf{D}$ is an $\mathrm{k} \times \mathrm{n}$ matrix. With rank factorization and singular value decomposition (SVD), any $\mathrm{n} \times \mathrm{n}$ matrix with rank $\mathrm{k}$ can be expressed as the products of some $\mathbf{B}$, $\mathbf{C}$, and $\mathbf{D}$. Thus, instead of an inversion of the $\mathrm{n} \times \mathrm{n}$ matrix, the inverse of the updated matrix can be computed by inverting a smaller matrix followed by matrix multiplications. Also, this equation does not involve any approximations, so it will provide the same result (in exact math) as directly inverting the updated matrix $\mathrm{n} \times \mathrm{n}$. Note that in order for this equation to be valid, we need to ensure that all $\mathbf{B}$, $\mathbf{C}$, and $\mathbf{D}$ matrices have rank $\mathrm{k}$ and adding the product of these matrices to the original $\mathbf{A}$ matrix does not result in zero determinant. 

Applying this to our Maxwell operator $\mathbf{A}$, with $\mathbf{B} = \mathbf{D}^\mathsf{T}=\mathbf{G}\mathbf{U}$ and $\mathbf{C} = \mathbf{diag}(-k_0^2 \mathbf{\Delta} {\epsilon})$, we have:

\begin{equation}
\begin{gathered}
    \mathbf{A'}^{-1} = \mathbf{A}^{-1} -\mathbf{A}^{-1}(\mathbf{G U}) (\mathbf{diag}(-k_0^2\mathbf{\Delta}{\epsilon})^{-1}+(\mathbf{GU})^\mathsf{T}\mathbf{A}^{-1}(\mathbf{G U}))^{-1} (\mathbf{GU})^\mathsf{T} \mathbf{A}^{-1}   \\
    = \mathbf{A}^{-1} -\mathbf{A}^{-1}\mathbf{G U} (\mathbf{diag}(-k_0^2\mathbf{\Delta}{\epsilon})^{-1}+\mathbf{U}^\mathsf{T}\mathbf{G}^\mathsf{T}\mathbf{A}^{-1}\mathbf{G U})^{-1} \mathbf{U}^\mathsf{T}\mathbf{G}^\mathsf{T}\mathbf{A}^{-1}
\end{gathered}   
\end{equation}

\section{Matrix Rank Update of the Precomputed Numerical Green Functions}
\label{SI:update}
Every time the physical structure is changed during the optimization, the current $\mathbf{Z}$, $\mathbf{Z_{obj}}$, and $\mathbf{E_{inc}}$ matrices are no longer correct for the updated structure. However, instead of rerunning the precomputation stage, the matrix rank update method can be applied to update the PNGFs as well. Thus, the computationally expensive precomputation stage only needs to be run once, after the design region and objective function have been defined, and the Green functions of any randomly initialized structure can then be computed with a few matrix rank updates to the PNGFs.

We recall that the three PNGFs are defined through the inverse of the system matrix $\mathbf{A}$ as
\begin{equation}
    \mathbf{Z} = \mathbf{G}^\mathsf{T}\mathbf{A}^{-1}\mathbf{G}, \qquad
    \mathbf{Z_{obj}} = \mathbf{S}^\mathsf{T}\mathbf{A}^{-1}\mathbf{G}, \qquad
    \mathbf{E_{inc}} = \mathbf{G}^\mathsf{T}\mathbf{A}^{-1}\mathbf{B}.
    \label{SI:def}
\end{equation}
When the physical structure is modified, the updated PNGFs are obtained simply by replacing $\mathbf{A}^{-1}$ with the perturbed inverse $\mathbf{A}'^{-1}$ in Eq.~(\ref{SI:def}). Substituting the expression for $\mathbf{A}'^{-1}$ derived in the previous section and recognizing that the recurring factor $\mathbf{G}^\mathsf{T}\mathbf{A}^{-1}\mathbf{G}$ inside the inner inverse is itself $\mathbf{Z}$, we obtain for $\mathbf{E_{inc}}$
\begin{align}
    \mathbf{E_{inc}'} 
    &= \mathbf{G}^\mathsf{T}\mathbf{A}'^{-1}\mathbf{B} \nonumber\\
    &= \mathbf{G}^\mathsf{T}\mathbf{A}^{-1}\mathbf{B}
       - \mathbf{G}^\mathsf{T}\mathbf{A}^{-1}\mathbf{G}\mathbf{U}
       \left(\mathbf{diag}(-k_0^2\mathbf{\Delta}\boldsymbol{\epsilon})^{-1}+\mathbf{U}^\mathsf{T}\mathbf{G}^\mathsf{T}\mathbf{A}^{-1}\mathbf{G}\mathbf{U}\right)^{-1}
       \mathbf{U}^\mathsf{T}\mathbf{G}^\mathsf{T}\mathbf{A}^{-1}\mathbf{B} \nonumber\\
    &= \mathbf{E_{inc}} - \mathbf{Z}\mathbf{U}
       \left(\mathbf{diag}(-k_0^2\mathbf{\Delta}\boldsymbol{\epsilon})^{-1}+\mathbf{U}^\mathsf{T}\mathbf{Z}\mathbf{U}\right)^{-1}
       \mathbf{U}^\mathsf{T}\mathbf{E_{inc}},
    \label{S6}
\end{align}
where we used $\mathbf{G}^\mathsf{T}\mathbf{A}^{-1}\mathbf{B}=\mathbf{E_{inc}}$ and $\mathbf{G}^\mathsf{T}\mathbf{A}^{-1}\mathbf{G}=\mathbf{Z}$. Applying the same substitution to the definitions of $\mathbf{Z_{obj}}$ and $\mathbf{Z}$ yields
\begin{align}
    \mathbf{Z_{obj}'} 
    &= \mathbf{S}^\mathsf{T}\mathbf{A}^{-1}\mathbf{G}
       - \mathbf{S}^\mathsf{T}\mathbf{A}^{-1}\mathbf{G}\mathbf{U}
       \left(\mathbf{diag}(-k_0^2\mathbf{\Delta}\boldsymbol{\epsilon})^{-1}+\mathbf{U}^\mathsf{T}\mathbf{Z}\mathbf{U}\right)^{-1}
       \mathbf{U}^\mathsf{T}\mathbf{G}^\mathsf{T}\mathbf{A}^{-1}\mathbf{G} \nonumber\\
    &= \mathbf{Z_{obj}} - \mathbf{Z_{obj}}\mathbf{U}
       \left(\mathbf{diag}(-k_0^2\mathbf{\Delta}\boldsymbol{\epsilon})^{-1}+\mathbf{U}^\mathsf{T}\mathbf{Z}\mathbf{U}\right)^{-1}
       \mathbf{U}^\mathsf{T}\mathbf{Z},
    \label{S7}
\end{align}
\begin{align}
    \mathbf{Z'} 
    &= \mathbf{G}^\mathsf{T}\mathbf{A}^{-1}\mathbf{G}
       - \mathbf{G}^\mathsf{T}\mathbf{A}^{-1}\mathbf{G}\mathbf{U}
       \left(\mathbf{diag}(-k_0^2\mathbf{\Delta}\boldsymbol{\epsilon})^{-1}+\mathbf{U}^\mathsf{T}\mathbf{Z}\mathbf{U}\right)^{-1}
       \mathbf{U}^\mathsf{T}\mathbf{G}^\mathsf{T}\mathbf{A}^{-1}\mathbf{G} \nonumber\\
    &= \mathbf{Z} - \mathbf{Z}\mathbf{U}
       \left(\mathbf{diag}(-k_0^2\mathbf{\Delta}\boldsymbol{\epsilon})^{-1}+\mathbf{U}^\mathsf{T}\mathbf{Z}\mathbf{U}\right)^{-1}
       \mathbf{U}^\mathsf{T}\mathbf{Z}.
    \label{S8}
\end{align}
In every case, each occurrence of the full inverse $\mathbf{A}^{-1}$ collapses into one of the precomputed matrices $\mathbf{Z}$, $\mathbf{Z_{obj}}$, or $\mathbf{E_{inc}}$.
In practice, Eqs.~(\ref{S6}) and (\ref{S7}) should be computed before Eq.~(\ref{S8}), since the $\mathbf{Z}$ matrix is overwritten by $\mathbf{Z'}$ in Eq.~(\ref{S8}).
\section{Computational Complexity Analysis}
\label{SI:computation_cost}
When evaluating the new structure after a tile flip during the optimization, we calculate the following equation using the precomputed matrices: 

\begin{equation}
\label{eq:obj}
    \Delta\mathbf{f} = -\mathbf{Z_{obj}}\mathbf{U} (\mathbf{diag}(-k_0^2\mathbf{\Delta}\boldsymbol{\epsilon})^{-1}+\mathbf{U}^\mathsf{T}\mathbf{Z}\mathbf{U})^{-1} \mathbf{U}^\mathsf{T}\mathbf{E_{inc}}
\end{equation}

In this equation, a careful grouping of matrices and ordering of computation is crucial to ensuring the high efficiency of the evaluation process. The complexity depends on system parameters: $\mathrm{N_{des}}$  (design region size), $\mathrm{N_{obj}}$ (number of objective functions), $\mathrm{N_{inc}}$ (number of excitations), $\mathrm{M}$ (size of a tile). For the device considered in this paper and most waveguide-based devices that can be optimized using the DBS algorithm, we can generally assume that $\mathrm{N_{obj}}, \mathrm{N_{inc}}, \mathrm{M} \ll \mathrm{N_{des}}$. In particular, the numbers of excitation sources and objective quantities, $\mathrm{N_{inc}}$ and $\mathrm{N_{obj}}$, respectively, are typically much smaller than both the tile size $\mathrm{M}$ and the number of design parameters $\mathrm{N_{des}}$.

Specifying the size of the matrices in this equation, $\mathbf{Z_{obj}}$ is a $\mathrm{N_{obj}} \times \mathrm{N_{des}}$ dense matrix; $\mathbf{U}$ is a $\mathrm{N_{des}} \times \mathrm{M}$ sparse matrix with $\mathrm{M}$ non-zero entries; $\mathbf{Z}$ is a $\mathrm{N_{des}} \times \mathrm{N_{des}}$ dense matrix; $\mathbf{E_{inc}}$ is a $\mathrm{N_{des}} \times \mathrm{N_{inc}}$ dense matrix. 

To calculate the $\mathrm{M}\times\mathrm{M}$ dense matrix $\mathbf{C} = (\mathbf{diag}(-k_0^2\mathbf{\Delta}\boldsymbol{\epsilon})^{-1}+\mathbf{U}^\mathsf{T}\mathbf{Z}\mathbf{U})$, we first calculate $\mathbf{Z}\mathbf{U}$, which requires $\mathrm{N_{des}M}$ operations. Then, we calculate $\mathbf{U}^\mathsf{T}(\mathbf{Z}\mathbf{U})$, where $\mathbf{Z}\mathbf{U}$ is now an $\mathrm{N_{des}}\times \mathrm{M}$ dense matrix, and perform the diagonal addition. The total number of operations required to compute the $\mathbf{C}$ matrix is thus ($\mathrm{M} + \mathrm{M^2} + \mathrm{N_{des}M}$). We further group and calculate $\mathbf{L} = \mathbf{Z_{obj}U}$ ($\mathrm{N_{obj}M}$ operations) and $\mathbf{R} = \mathbf{\mathbf{U}^\mathsf{T}E_{inc}}$ ($\mathrm{N_{inc}M}$ operations).
Since solving the system of equations $\mathbf{C}^{-1}\mathbf{R}$ requires $\mathrm{O(M^3+M^2N_{inc})}$ operations, the total complexity of the evaluation is $\mathrm{M(N_{des}+N_{obj}+N_{inc}+N_{obj}N_{inc}+M) + O(M^3+M^2N_{inc})}$. With the assumptions stated above, this reduces to $\mathrm{O(N_{des})}$, i.e., linear scaling with respect to the size of the design region.

When updating the precomputed matrices, changes in the $\mathbf{Z}$ matrix dominate the computational complexity: 

\begin{equation}
    \Delta\mathbf{Z} = -\mathbf{Z}\mathbf{U} (\mathbf{diag}(-k_0^2\mathbf{\Delta}\boldsymbol{\epsilon})^{-1}+\mathbf{U}^\mathsf{T}\mathbf{Z}\mathbf{U})^{-1} \mathbf{U}^\mathsf{T}\mathbf{Z}
\end{equation}

Following a similar grouping, we can reuse the $\mathrm{M}\times\mathrm{M}$ dense matrix $\mathbf{C}$ and the $\mathrm{N_{des}}\times\mathrm{M}$ dense matrix $\mathbf{L'} = \mathbf{ZU}$. Computing $\mathbf{R'} = \mathbf{U}^\mathsf{T}\mathbf{Z}$ requires $\mathrm{N_{des}M}$ operations, and solving the system of equations $\mathbf{C}^{-1}\mathbf{R'}$ again requires $\mathrm{O(M^3+M^2N_{des})}$ operations. In this case, the multiplication $\mathbf{L'}(\mathbf{C}^{-1}\mathbf{R'})$ dominates the computational complexity, because both matrices are dense, with sizes $\mathrm{N_{des}}\times\mathrm{M}$ and $\mathrm{M}\times\mathrm{N_{des}}$, respectively. Overall, the complexity of updating the system matrices scales as $\mathrm{O(N_{des}^2)}$.

\section{Mode Overlap Integral}
\label{SI:mode_overlap}
Nanophotonic design often requires evaluation of the power transmitted in certain modes in both input and output waveguides. The propagation of electromagnetic fields along an infinite waveguide can be expanded into superpositions of a set of known waveguide modes propagating in both forward and backward directions as: 
\begin{align}
    \mathbf{E} &= \sum_m \left( a_m \mathbf{E}_{m}^{\mathrm{fwd}} + b_m \mathbf{E}_{m}^{\mathrm{bwd}} \right) \\
    \mathbf{H} &= \sum_m \left( a_m \mathbf{H}_{m}^{\mathrm{fwd}} + b_m \mathbf{H}_{m}^{\mathrm{bwd}} \right)
\end{align}
where $a_m$ and $b_m$ are the complex amplitudes of the forward and backward propagating modes respectively. The target mode profiles can be obtained by solving the waveguide eigenmode problem on the cross section. For a structure that is invariant along the propagation direction, the field dependence can be written as $\exp(-\mathrm{i}\beta_m z)$, 
where $\beta_m$ is the propagation constant of the mode. Once $\mathbf{E}_{m}^{\mathrm{fwd}}$ is solved, the corresponding magnetic field can be obtained from Maxwell's equation
\begin{equation}
    \mathbf{H}_{m}^{\mathrm{fwd}}
    =
    -\frac{1}{\mathrm{i}\omega\mu}
    \nabla\times \mathbf{E}_{m}^{\mathrm{fwd}} .
\end{equation}
In numerical implementations, the eigenmode problem can be discretized on the same Yee grid as the full Maxwell simulation. Therefore, the modal fields can be sampled or interpolated onto the same electric-field and magnetic-field locations used in the design simulation.
Using the orthogonality relation of waveguide modes, the amplitude of a desired forward output mode can be extracted through an overlap integral between the simulated total field and the target forward waveguide mode profile:
\begin{equation}
    a_m =
    \frac{1}{P_m}
    \int_A
    \left(
    \mathbf{E} \times
    \mathbf{H}_{m}^{\mathrm{fwd},*}
    \right)\cdot \hat{\mathbf{n}}\, dA ,
\end{equation}
where $\mathbf{E}$ is the simulated electric field, $\mathbf{H}_{m}^{\mathrm{fwd}}$ is the magnetic field profile of the target forward mode, $\hat{\mathbf{n}}$ is the propagation direction normal to the cross section, and
\begin{equation}
    P_m =
    \int_A
    \left(
    \mathbf{E}_{m}^{\mathrm{fwd}} \times
    \mathbf{H}_{m}^{\mathrm{fwd},*}
    \right)\cdot \hat{\mathbf{n}}\, dA
\end{equation}
is the normalization constant corresponding to the modal power. After normalization, the transmitted power coupled into the target mode is
\begin{equation}
    T_m = |a_m|^2 .
\end{equation}
In the discretized formulation used in this work, this overlap operation can be represented as a linear projection of the total field vector onto the target mode profile. This projection is represented by the matrix $\mathbf{S}$ in the main text. Each column of $\mathbf{S}$ contains the discretized overlap kernel corresponding to one objective mode profile. For an electric-field-only formulation, this kernel is primarily constructed from the conjugated target modal magnetic field $\mathbf{H}_{m}^{\mathrm{fwd},*}$, with the proper Yee-grid interpolation and quadrature weights. Therefore,
\begin{equation}
    \mathbf{Z}_{\mathrm{obj}} = \mathbf{S}^{\mathsf{T}}\mathbf{A}^{-1}\mathbf{G}
\end{equation}
maps equivalent currents inside the design region to the complex modal amplitudes associated with the chosen objective functions. The magnitude squared of these complex overlap coefficients corresponds to the optical power coupled into the desired output modes.

\section{Modified Direct Binary Search (DBS) for Improved Connectivity}
\label{SI:modified_DBS}
Although the standard direct binary search (DBS) algorithm can achieve high efficiency, it often yields QR-code-like structures with numerous isolated silicon ``islands'' and air ``holes,'' which are undesirable from a fabrication standpoint. To incorporate connectivity constraints into the optimization, we modify DBS by combining a connectivity-aware search strategy with a post-processing clean-up stage. The detailed procedures of the standard and modified DBS algorithms are summarized in Algorithms~\ref{alg:dbs} and~\ref{alg:modified}, respectively.

The design is initialized as an all-silicon structure ($x_i = 1$), after which a small number of air pixels ($x_i = 0$) are randomly introduced to ensure variability between optimization trials. During the optimization, pixel updates are prioritized based on their proximity to existing air regions. Specifically, pixels adjacent to existing holes are evaluated first, promoting the growth and merging of connected void regions. Only when no improving flips are found near existing holes does the algorithm consider flipping pixels in the background region, allowing the creation of new holes or the filling of existing ones.

After the DBS procedure converges to a local optimum (i.e., no single-pixel flip improves the objective), a clean-up stage is applied. In this stage, each pixel is examined to determine whether it is isolated with respect to its neighbors. If a pixel is identified as isolated, it is flipped and the objective is reevaluated. Unlike the main optimization stage, such flips are accepted if the resulting degradation in the objective function is within a prescribed tolerance. This allows removal of isolated features while maintaining overall device performance.

\begin{algorithm}[H]
    \captionof{algorithm}{Direct Binary Search (DBS)}
    \label{alg:dbs}
    \begin{algorithmic}[1]
        \State Initialize binary design $\mathbf{x} \in \{0,1\}^N$
        \State Evaluate objective $f(\mathbf{x})$
        \State Mark all pixels as \textbf{unchecked}

        \While{true}
            \If{all pixels are checked}
                \State \textbf{break} \Comment{no improving single-tile flip exists}
            \EndIf

            \State Randomly select an unchecked pixel $x_i$
            \State Flip pixel: $x_i \leftarrow 1 - x_i$
            \State Evaluate $f_{\text{new}}$

            \If{$f_{\text{new}} > f(\mathbf{x})$}
                \State Accept change: $f(\mathbf{x}) \leftarrow f_{\text{new}}$
                \State Mark all pixels as \textbf{unchecked} \Comment{restart search}
            \Else
                \State Reject change: $x_i \leftarrow 1 - x_i$
                \State Mark pixel $x_i$ as \textbf{checked}
            \EndIf
        \EndWhile

        \State \Return $\mathbf{x}$
    \end{algorithmic}
\end{algorithm}

\begin{algorithm}[H]
    \captionof{algorithm}{Modified Direct Binary Search with Connectivity Preference and Clean-Up}
    \label{alg:modified}
    \begin{algorithmic}[1]
        \State Initialize $\mathbf{x} \in \{0,1\}^N$ with $x_i = 1$ (silicon) for all $i$
        \State Randomly set a subset of pixels to $x_i = 0$ (air holes)
        \State Evaluate objective $f(\mathbf{x})$
        \State Mark all pixels as \textbf{unchecked}

        \State Partition pixels:
        \Statex \quad $\mathcal{G}_1$: pixels adjacent to at least one $x=0$ pixel
        \Statex \quad $\mathcal{G}_2$: all remaining pixels

        \While{true}
            \If{all pixels are checked}
                \State \textbf{break} \Comment{no improving single-pixel flip exists}
            \EndIf

            \If{$\mathcal{G}_1 \neq \emptyset$}
                \State Randomly select an unchecked pixel $x_i \in \mathcal{G}_1$
            \Else
                \State Randomly select an unchecked pixel $x_i \in \mathcal{G}_2$
            \EndIf

            \State Flip pixel: $x_i \leftarrow 1 - x_i$
            \State Evaluate $f_{\mathrm{new}}$

            \If{$f_{\mathrm{new}} > f(\mathbf{x})$}
                \State Accept flip: $f(\mathbf{x}) \leftarrow f_{\mathrm{new}}$
                \State Mark all pixels as \textbf{unchecked} \Comment{restart search}

                \State Repartition pixel groups $\mathcal{G}_1$ and $\mathcal{G}_2$
            \Else
                \State Reject flip: $x_i \leftarrow 1 - x_i$
                \State Mark pixel $x_i$ as \textbf{checked}
                \State Remove element corresponding to $x_i$ in selected group
            \EndIf
        \EndWhile

        \State \Comment{Clean-up stage}
        \For{each pixel $i$}
            \If{$x_i$ is isolated with respect to its neighbors}
                \State Flip pixel: $x_i \leftarrow 1 - x_i$
                \State Evaluate $f_{\mathrm{new}}$

                \If{$f_{\mathrm{new}} \geq f(\mathbf{x}) - \delta$}
                    \State Accept clean-up flip: $f(\mathbf{x}) \leftarrow f_{\mathrm{new}}$
                \Else
                    \State Reject flip: $x_i \leftarrow 1 - x_i$
                \EndIf
            \EndIf
        \EndFor

        \State \Return $\mathbf{x}$
    \end{algorithmic}
\end{algorithm}

\section{Level-set parameterization}
\label{SI:level_set}
\subsection{Radial Basis Functions (RBF) for level-set Function Construction}

The level-set function can be constructed as a sum of RBFs centered at knots placed (usually uniformly) on a coarse grid:
\begin{equation}
\label{eq:RBF1}
    {\phi(\mathbf{r})} = {\sum_{i=1}^N \alpha_i \psi(\mathbf{\|r-r_i\|})}, 
\end{equation}
where
\begin{itemize}
    \item $\mathbf{r}$: evaluation point
    \item $\mathbf{r}_i$: RBF centers (nodes)
    \item $\alpha_i$: control coefficients
    \item $\psi(r)$: the RBF, chosen in this work as the Gaussian basis function $e^{-\frac{r^2}{2\sigma^2}}$
    \item $\|\cdot\|$: Euclidean distance
\end{itemize}

The spread parameter $\sigma$ is chosen so that there is good overlap between neighboring knots, preventing ill-conditioned systems or overly smoothed designs. In this work, $\sigma$ is chosen to be 2 or 3 times the knot-grid resolution. Eq. (\ref{eq:RBF1}) can also be written in the more compact discretized matrix form $\Phi_{N_{ls}\times 1} = \Psi_{N_{ls}\times N_{\rho}} \times \boldsymbol{\alpha}_{N_{\rho} \times 1}$, where $N_{\rho}$ and $N_{ls}$ are the numbers of pixels on the knot grid and the level-set grid, respectively.

\subsection{Single-Knot Update with Cutoff Radius}
Instead of updating the entire design region, only one control coefficient $\alpha_k$ is updated at a time:
\begin{equation}
\phi^{\,new}(\mathbf{x})=\phi^{\,old}(\mathbf{r})+\Delta
 \alpha_k \,\psi\!\left(\frac{\|\mathbf{r}-\mathbf{r}_k\|}{r_c}\right),
\qquad \|\mathbf{r}-\mathbf{r}_k\|\le r_c
\end{equation}
where $r_c$ is the cutoff radius (in this work, 3 or 4 times the knot-grid resolution), so that:

\begin{equation}
\psi_c(r) =
\begin{cases}
\psi(r), & r \le r_c \\
0, & r > r_c
\end{cases}
\end{equation}
with $r = \mathbf{\|r-r_k\|}$. Therefore, only the subset of pixels belonging to $B(\mathbf{r}_k, r_c) = \left\{ \mathbf{r} : \|\mathbf{r} - \mathbf{r}_k\| \le r_c \right\}$ is updated. In other words, the update only affects a ball centered at the updated control knot.

\subsection{Gradient with Respect to the Parameterization}

Since $ {\epsilon_{ls}\mathbf{(x,y)}}= \epsilon_1 + \frac{1}{2}(\epsilon_2-\epsilon_1) [ \tanh({\beta {\phi\mathbf{(x,y)}}) }+1 ]$ we have:
\begin{equation}
\label{eq:dsdls}
\frac{\partial \epsilon^i_{ls}}{ \partial \phi^i} = \frac{\beta (\epsilon_2- \epsilon_1)}{2} \mathrm{sech}^2 \! [\beta { \phi^i\mathbf{(x,y)}} ],
\end{equation}
where $i$ denotes the pixel being evaluated. With the cutoff radius introduced, Eq. (\ref{eq:dsdls}) can be updated element-wise as well, with only the values corresponding to the subset $B(\mathbf{r}_k, r_c)$ being updated instead of re-evaluating the whole design region.

Overall, the gradient of the permittivity profile on the level-set grid with respect to the control knots is:

\begin{equation}
\label{eq:grad1}
    \frac{\partial \epsilon_{ls}}{ \partial \alpha} = \left[\frac{\partial \epsilon_{ls}}{ \partial \phi}\right]_{N_{ls}\times 1} \odot \Psi_{N_{ls}\times N_{\rho}}, 
\end{equation}
with $\odot$ denoting the Hadamard (element-wise) product (in this case, can be understood as element-wise product of matrix and the implicit expansion of vector). This can also be calculated efficiently, since each level-set point belongs to only a small number of subsets $B(\mathbf{r}_k, r_c)$.

\subsection{Volumetric Averaging}
For a coarser simulation cell, the downsampled or subpixel-averaged permittivity is:
\begin{equation}
\bar{\epsilon}_{sim} = \frac{1}{|V|} \int_V \epsilon(\mathbf{r}) \, d\mathbf{r},
\end{equation}
where $V$ denotes the volume/area of the simulation cell. This can be rewritten as the matrix multiplication $\left[\bar{\epsilon}_{sim}\right]_{N_{sim} \times 1} = [D]_{N_{sim}\times N_{ls}}\times \left[\epsilon_{ls} \right]_{N_{ls} \times 1}$; therefore, $\partial \bar{\epsilon}_{sim}/ \partial \epsilon_{ls} = D$. The final derivative $\partial \bar{\epsilon}_{sim}/ \partial \alpha$ is calculated by downsampling $\partial \epsilon_{ls} / \partial \alpha$ obtained from Eq. (\ref{eq:grad1}), which can also be updated element-wise efficiently, since each $\epsilon_{ls}$ value belongs to only a few simulation cells.

\section{Additional Inverse Design Examples using DBS}
\label{SI:examples}
\subsection{Waveguide Bend}
\label{sec:WGB_DBS}
\begin{figure}[ht]
\centering
\fbox{\includegraphics[width=13cm]{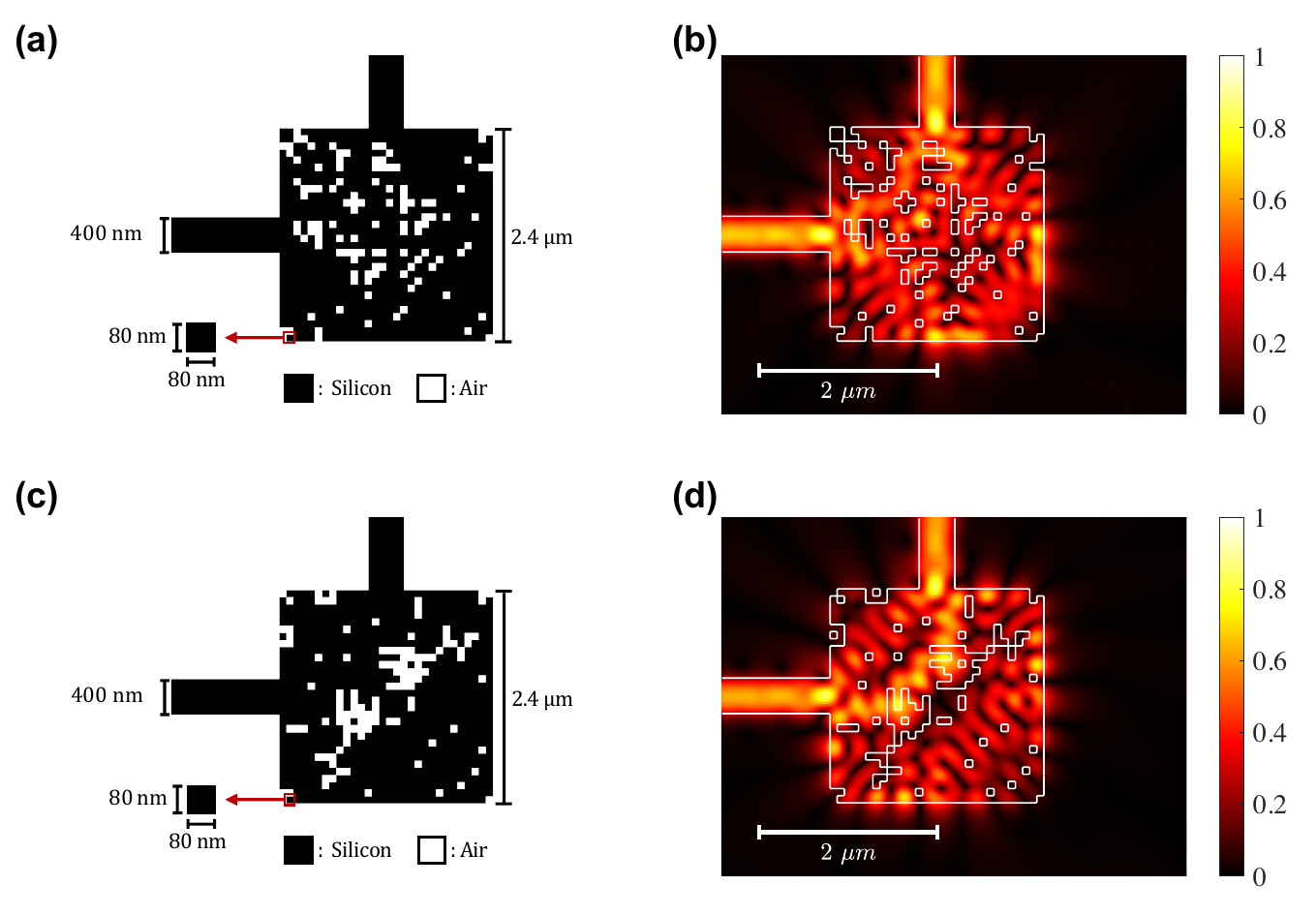}}
\caption{Waveguide bends inverse designed using DBS. (a) Physical structure of the design optimized with standard DBS. (b) Normalized $|E_z|$ distribution of the structure in (a). (c) Physical structure of the design optimized with the modified (connectivity-enhanced) DBS. (d) Normalized $|E_z|$ distribution of the structure in (c).}
\label{fig:WGB_DBS}
\end{figure}

We next consider a waveguide bend that redirects the input power, which is initially in the fundamental mode, into a direction perpendicular to its original propagation. The goal is to maximize coupling into the fundamental mode of the output waveguide while minimizing reflection and radiation losses. The objective function in this case is simply the mode overlap between the target mode and the field profile at the output of the waveguide bend. The design region has dimensions of $2.4 \times 2.4~\mu\text{m}^2$, and the input/output waveguides are $0.4~\mu\text{m}$ wide. Each tile occupies a physical area of $80 \times 80~\text{nm}^2$, and mirror symmetry is enforced about the $45^{\circ}$ axis, as shown in Fig. \ref{fig:WGB_DBS}. The final design achieves a transmission efficiency of up to $99.3\%$ after 102 good flips and 5604 bad flips in 17 seconds, indicating that nearly all of the input power in the fundamental mode is successfully redirected. For the modified DBS, the optimization converged after 130 good flips and 5816 bad flips in 24.7 seconds, with a total transmission of $\approx 99.5\%$.

\begin{table}[H]
\centering
\caption{\bf Convergence and Acceleration of Waveguide Bend Optimization using DBS}
\label{table:WGB_DBS}
\resizebox{\textwidth}{!}{%
\begin{tabular}{ccccc}
\hline
 FDTD (Lumerical) & FDFD & PNGF & Speed Up (FDTD) & Speed Up (FDFD) \\
\hline
 6.34 h & 0.8 h & 16.92 seconds & 1348x & 170x\\
\hline
\end{tabular}
}
\end{table}

\subsection{Arbitrary Power Splitter}

\begin{figure}[H]
\centering
\fbox{\includegraphics[width=12cm]{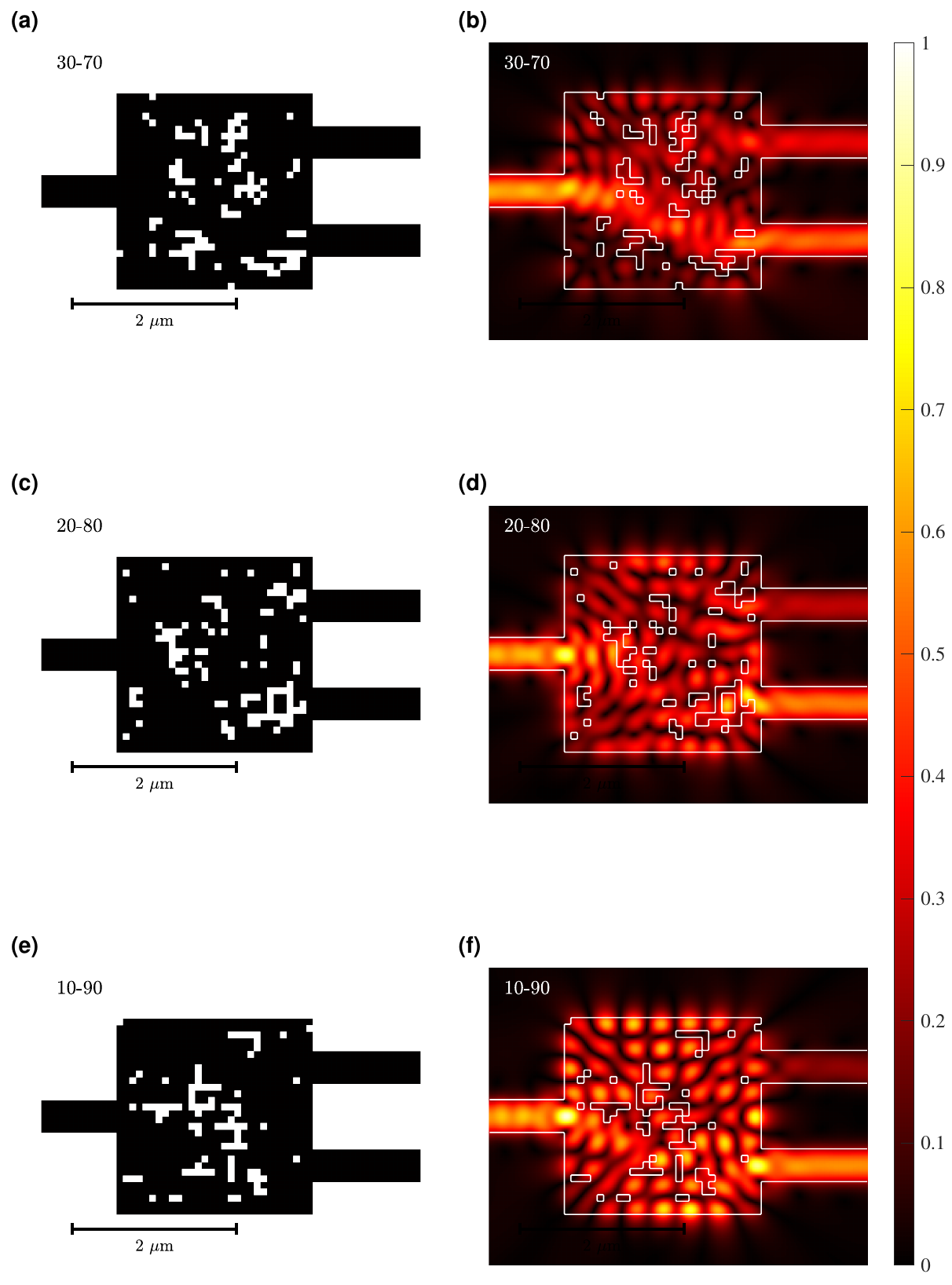}}
\caption{\label{fig:PS_DBS_arbi_C} Arbitrary-ratio power splitters optimized using the modified (connectivity-enhanced) DBS. Left column: optimized structures; right column: corresponding normalized $|E_z|$ distributions.}
\end{figure}

In some applications, an integrated power splitter with a customized splitting ratio may be preferred. In this case, however, symmetry can no longer be enforced when designing the device. With a setup similar to that of the 50:50 power splitter, we can even reuse the same precomputed matrix $\mathbf{Z}$. The final results are shown below: Fig. \ref{fig:PS_DBS_arbi} presents the structures obtained with standard DBS, and Fig. \ref{fig:PS_DBS_arbi_C} presents the structures obtained with connectivity taken into consideration. The acceleration achieved in the two cases is summarized in Tables \ref{SI-table:PS_DBS} and \ref{table:PS_DBS_C}.

\begin{table}[h]
\centering
\caption{\bf Convergence and Acceleration of Power Splitter using DBS}
\label{SI-table:PS_DBS}
\resizebox{\textwidth}{!}{%
\begin{tabular}{cccccc}
\hline
 Ratio & FDTD (Lumerical) & FDFD & PNGF & Speed Up (FDTD) & Speed Up (FDFD) \\
\hline
50-50 & 6.36 h & 0.8 h & 12.1 seconds & 1579$\times$ & 197$\times$\\
\hline
70-30 & 36.5 h & 4.7 h & 21.7 seconds & 6055$\times$ & 779$\times$\\
\hline
80-20 & 32.0 h & 4.2 h & 24.9 seconds & 4626$\times$ & 607$\times$\\
\hline
90-10 & 27.5 h & 3.6 h & 19.1 seconds & 5183$\times$ & 678$\times$\\
\hline
\end{tabular}
}
\end{table}

\begin{figure}[H]
\centering
\fbox{\includegraphics[width=12cm]{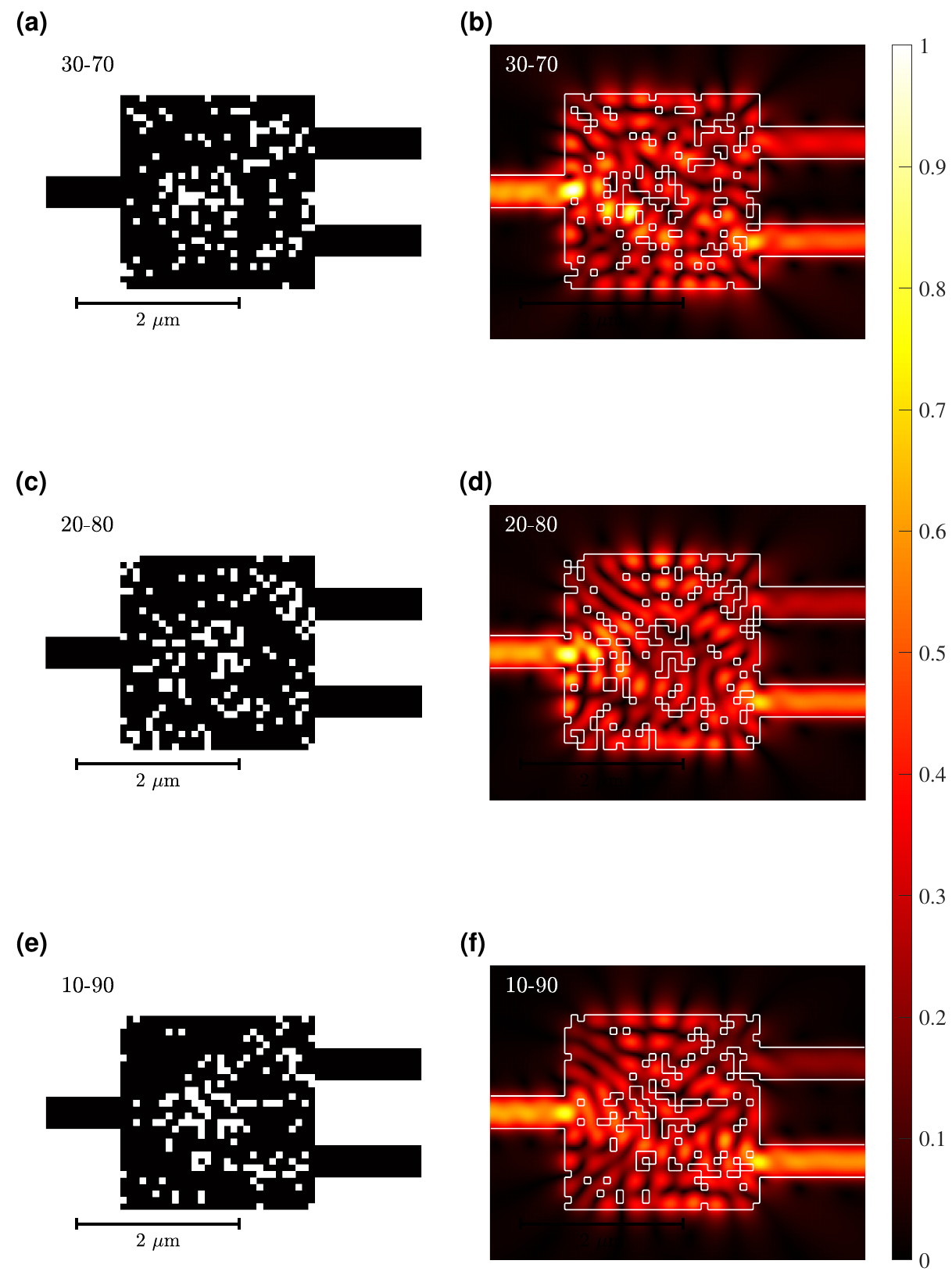}}
\caption{\label{fig:PS_DBS_arbi} Arbitrary-ratio power splitters optimized using standard DBS for target splitting ratios of 30:70 (a,b), 20:80 (c,d), and 10:90 (e,f). Left column: optimized structures; right column: corresponding normalized $|E_z|$ distributions.}
\end{figure}

\begin{table}[H]
\centering
\caption{\bf Convergence and Acceleration of Power Splitter using Modified DBS}
\label{table:PS_DBS_C}
\resizebox{\textwidth}{!}{%
\begin{tabular}{cccccc}
\hline
 Ratio & FDTD (Lumerical) & FDFD & PNGF & Speed Up (FDTD) & Speed Up (FDFD) \\
\hline
50-50 & 9.5 h & 1.2 h & 12.2 seconds & 2818$\times$ & 354$\times$\\
\hline
70-30 & 35.6 h & 4.6 h & 22.9 seconds & 5596$\times$ & 723$\times$\\
\hline
80-20 & 45.0 h & 5.9 h & 22.1 seconds & 7330$\times$ & 961$\times$\\
\hline
90-10 & 55.5 h & 7.2 h & 29.5 seconds & 6772$\times$ & 881$\times$\\
\hline
\end{tabular}
}
\end{table}

\subsection{Mode Converter using Rotatable DBS}
\label{supSI:RDBS}
Unfortunately, ordinary DBS may become inefficient or fundamentally limited for devices that require precise phase engineering, e.g., phase-sensitive structures such as multimode multiplexers, wavelength demultiplexers, and polarization splitters. In conventional DBS, the design region is split into many rectangular tiles, which generate abrupt, strongly scattering perturbations and provide no means of phase control. Rotatable DBS (RDBS) \cite{Ma2020RDBS} was introduced to address this issue: the design region is split into cells, each holding an air hole of arbitrary shape that is rotated by an adjustable angle rather than simply switched on or off. This facilitates better phase alignment compared to standard DBS.

\begin{figure}[H]
\centering
\fbox{\includegraphics[width=13cm]{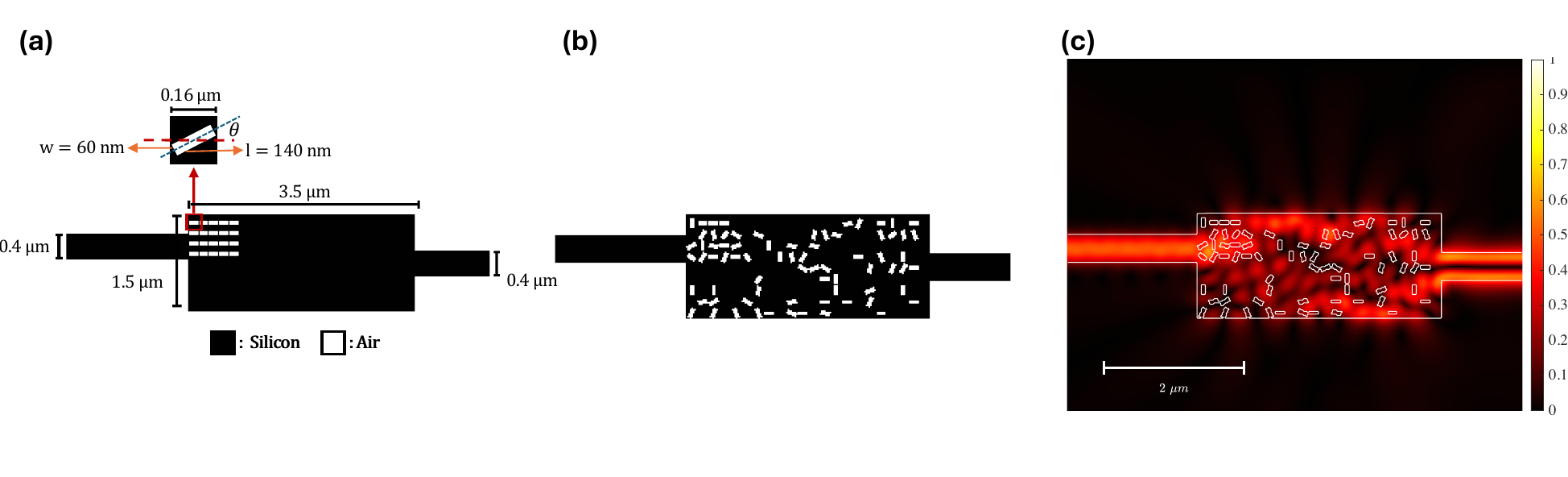}}
\caption{\label{fig:MC_RDBS} Mode converter inverse designed using rotatable DBS (RDBS). (a) Initial guess with the device dimensions and the unit cell description: each $0.16~\mu\text{m}$ cell holds a $60 \times 140~\text{nm}$ air slot rotated by an adjustable angle $\theta$. (b) Physical structure of the optimized design. (c) Normalized $|E_z|$ distribution of the final structure, showing conversion of the input $TM_0$ mode to the $TM_1$ mode at the output.}
\end{figure}

In this example, we combine RDBS with PNGF to inverse design a mode converter ($TM_0$ to $TM_1$). The dimensions of the device and a description of the unit cells are given in Fig. \ref{fig:MC_RDBS}. In the optimization loop, the algorithm sweeps across the structure cell by cell, rotating each unit cell through candidate angles $\theta$ in increments of $10^{\circ}$, selecting the best angle, and updating the structure accordingly. To penalize crosstalk and maximize the output in the target mode, the FoM is defined as:

\begin{equation}
    f_{FoM} = \frac{P_{TM_1}}{P_{TM_0}+0.01}\times \log_{10}{{(P_{TM_1}+1.1)}}
\end{equation}

Since $TM_0$ is a symmetric mode while $TM_1$ is antisymmetric, the input waveguide is positioned off-center relative to the junction to introduce asymmetry into the system. In addition, a cell array is introduced in the top-left corner of the structure to further enhance the asymmetry. The optimization took 15 sweeps over 190 cells (in random order), producing a structure with 97.5\% conversion efficiency in 36.1 seconds.

\begin{table}[H]
\centering
\caption{\bf Convergence and Acceleration of Mode Converter using RDBS}
\label{table:MC_RDBS}
\resizebox{\textwidth}{!}{%
\begin{tabular}{ccccc}
\hline
 FDTD (Lumerical) & FDFD & PNGF & Speed Up (FDTD) & Speed Up (FDFD) \\
\hline
 172.0 h & 28.4 h & 36.1 seconds & 17152$\times$ & 2832$\times$\\
\hline
\end{tabular}
}
\end{table}

\section{Additional Inverse Design Examples using the Level-Set Method}
\label{SI:examples_levelset}
\subsection{Waveguide Bend}
\begin{figure}[htbp]
\centering
\fbox{\includegraphics[width=13cm]{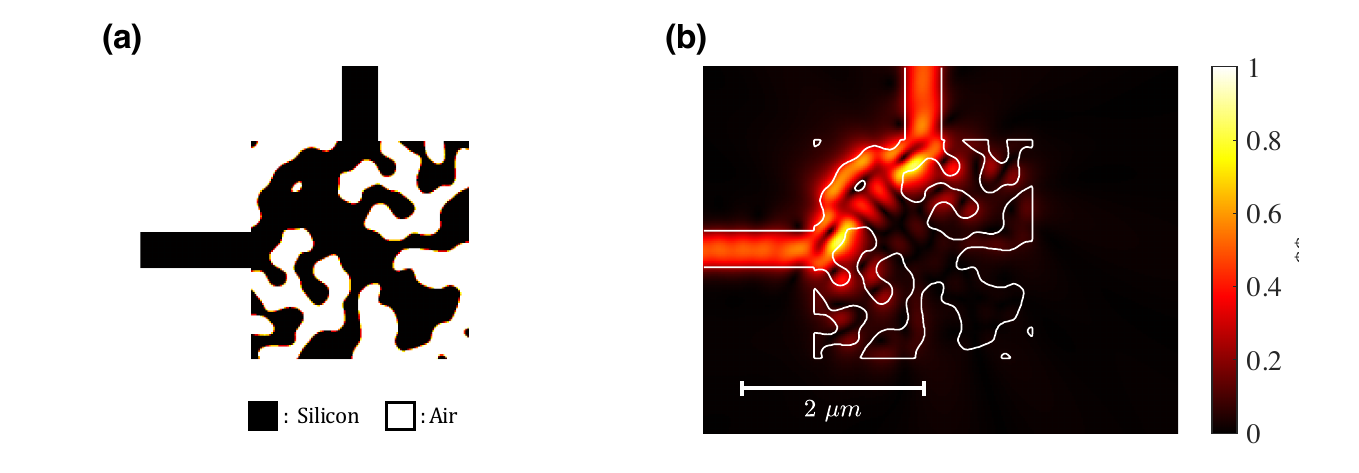}}
\caption{(a) Physical structure of the waveguide bend optimized using the level-set method. (b) Normalized $|E_z|$ distribution of the final structure.}
\label{fig:WGB_LS_rand}
\end{figure}

The same parameterization setup as for the power splitter is retained for inverse designing a $90^\circ$ waveguide bend, except that symmetry is enforced about the $45^\circ$ axis of the junction. The junction again has a $2.4 \times 2.4~\mu\text{m}^2$ footprint. As in Section \ref{sec:WGB_DBS}, we maximize the power delivered to the fundamental mode of the output waveguide. Figure \ref{fig:WGB_LS_rand} shows the final structure of the device and its performance at $1550~\text{nm}$. After 370 iterations, completed in 33.3 seconds, the device reached a performance of almost $99.5 \%$ of the input power delivered to the output port.

\begin{table}[H]
\centering
\caption{\bf Convergence and Acceleration of Waveguide Bend Optimization using the Level-Set Method}
\label{table:WGB_LS}
\resizebox{\textwidth}{!}{%
\begin{tabular}{ccccc}
\hline
 FDTD (Lumerical) & FDFD & PNGF & Speed Up (FDTD) & Speed Up (FDFD) \\
\hline
 2.3 h & 0.2 h & 33.3 seconds & 255$\times$ & 22$\times$\\
\hline
\end{tabular}
}
\end{table}

\subsection{Mode Converter}

\begin{figure}[H]
\centering
\fbox{\includegraphics[width=\linewidth]{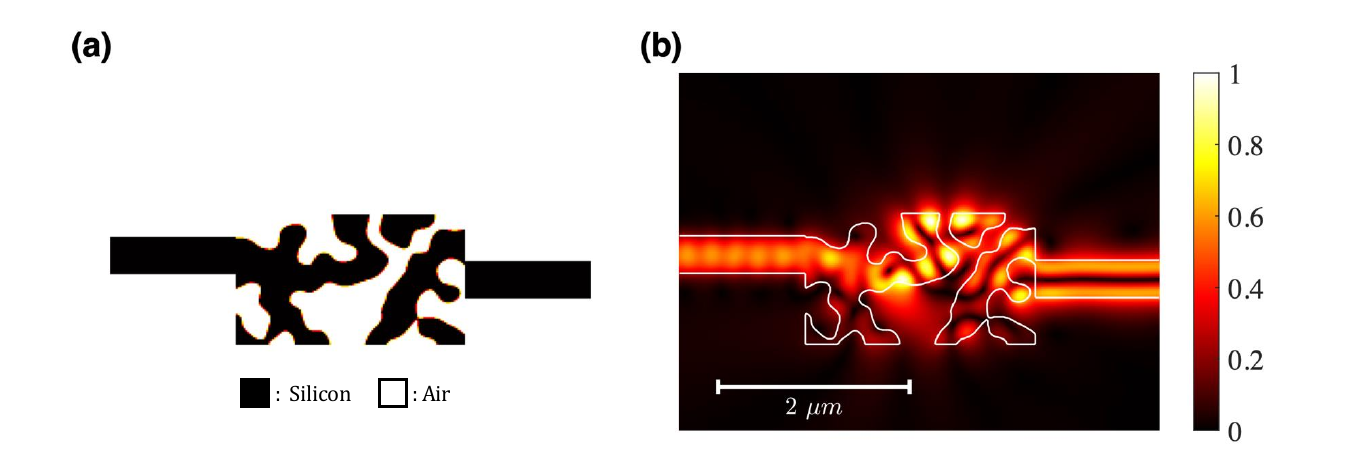}}
\caption{\label{fig:MC_LS_rand} (a) Physical structure of the mode converter optimized using the level-set method. (b) Normalized $|E_z|$ distribution of the final structure, showing conversion of the input $TM_0$ mode to the $TM_1$ mode at the output.}
\end{figure}

Although the RDBS approach introduced in Section \ref{SI:examples}\ref{supSI:RDBS} achieved reasonable performance, several limitations remain when designing more compact devices. Here, the level-set method proves more effective for designing the mode converter, as it allows continuous modification of the design. The objective function optimizes the transmitted modal amplitude of the output mode ($TM_1$) excited from the input mode ($TM_0$), and the final efficiency is then calculated using the mode overlap. The junction occupies a $1.4 \times 2.4~\mu\text{m}^2$ region, much smaller than in the RDBS case. The final structure, shown in Fig. \ref{fig:MC_LS_rand}, emerged after 730 iterations in 28.8 seconds, with a final conversion efficiency of $98.6\%$ calculated at the output.

\begin{table}[H]
\centering
\caption{\bf Convergence and Acceleration of Mode Converter Optimization using the Level-Set Method}
\label{table:MC_LS}
\resizebox{\textwidth}{!}{%
\begin{tabular}{ccccc}
\hline
 FDTD (Lumerical) & FDFD & PNGF & Speed Up (FDTD) & Speed Up (FDFD) \\
\hline
 3.6 h & 18 mins & 28.8 seconds & 389$\times$ & 37$\times$\\
\hline
\end{tabular}
}
\end{table}

\subsection{Arbitrary-Ratio Power Splitters with a Y-Junction Initial Guess}

\begin{figure}[H]
\centering
\fbox{\includegraphics[width=12cm]{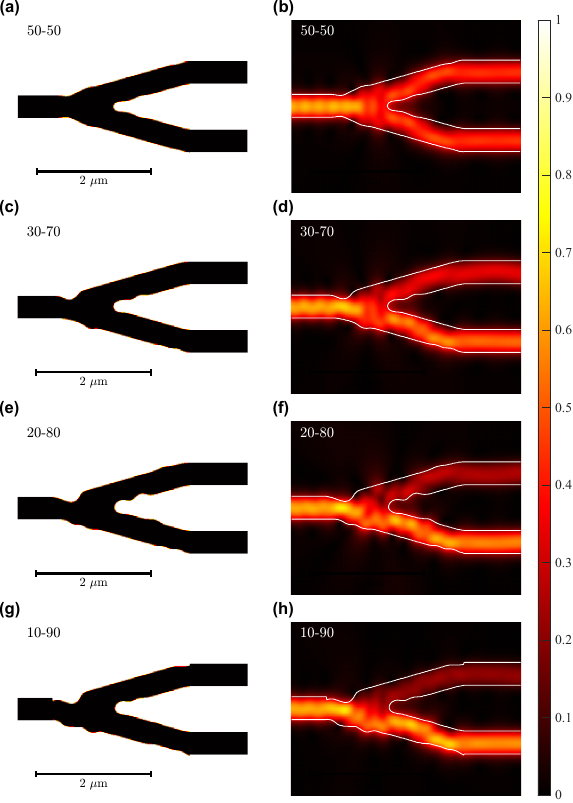}}
\caption{\label{fig:PS_ini_LS} Power splitters refined from a Y-junction initial guess using the level-set method, for target splitting ratios of 50:50 (a,b), 30:70 (c,d), 20:80 (e,f), and 10:90 (g,h). Left column: optimized structures; right column: corresponding normalized $|E_z|$ distributions.}
\end{figure}

Although the level-set method can work with a random initial guess, it also works particularly well as a refinement method. Instead of starting from scratch, a well-defined or already functional structure can be used as the initial guess, allowing the optimization to focus on improving the existing topology through smooth boundary evolution and local shape adjustment, which is especially suitable for fine-tuning devices. Here, we present optimizations initialized from a Y-branch structure for arbitrary-ratio power splitters. The structures have the same junction size as in the previous sections and evolve incrementally to the optimized designs shown in Fig. \ref{fig:PS_ini_LS}. As expected, the number of iterations and the computational time are reduced significantly compared with randomly initialized optimization (Table \ref{table:PS_LS}).

\begin{table}[H]
\centering
\caption{\bf Convergence and Acceleration of Power Splitter Structure Refinement using the Level-Set Method}
\label{table:PS_LS}
\resizebox{\textwidth}{!}{%
\begin{tabular}{cccccc}
\hline
 Ratio & FDTD (Lumerical) & FDFD & PNGF & Speed Up (FDTD) & Speed Up (FDFD) \\
\hline
50-50 & 28.8 minutes & 2.5 minutes & 9.7 seconds & 177$\times$ & 15$\times$\\
\hline
70-30 & 16.5 minutes & 1.5 minutes & 6.0 seconds & 164$\times$ & 15$\times$\\
\hline
80-20 & 17.3 minutes& 1.5 minutes & 6.7 seconds & 154$\times$ & 13$\times$\\
\hline
90-10 & 59.0 minutes & 5.1 minutes & 15.8 seconds & 224$\times$ & 19$\times$\\
\hline
\end{tabular}
}
\end{table}

\putbib[supp_info/sample] 
\end{bibunit}

\end{document}